\documentclass[11pt]{article}

\usepackage{epsfig,amsmath,latexsym,amssymb, amsthm}
\usepackage{graphicx}
\usepackage{lscape}
\usepackage{picture, eso-pic, tikz} 
\usepackage{dsfont}
\usepackage{listings}
\usepackage{changes}
\usepackage{hyperref}
\usepackage{bbding}
\usepackage{booktabs}

\renewcommand{\baselinestretch}{1.1}
\def\R{{\mathbb R}}  
\def\p{{\mathbb P}}  
\def\E{{\mathbb E}}  %
\def\Beweis{\footnotesize}

\newcommand{\Remm}[1]{}
\newtheorem{theo}{Theorem}[section]
\newtheorem{lemma}[theo]{Lemma}
\newtheorem{prop}[theo]{Proposition}

\newtheorem{cor}[theo]{Corollary}
\newtheorem{defi}[theo]{Definition}
\newtheorem{model ass}[theo]{Model Assumptions}

\newtheorem{example}[theo]{Example}

\newtheorem{rem}[theo]{Remark}
\def\EndProof{\hfill {\scriptsize $\Box$}}
\def\EndExample{\hfill {\scriptsize $\blacksquare$}}
\numberwithin{equation}{section}

\newcommand{\specialcell}[2][c]{%
\begin{tabular}[#1]{@{}l@{}}#2\end{tabular}}

\definecolor{MyGray}{rgb}{0.92,0.92,0.92}


\definecolor{British racing}{rgb}{0.0, 0.5, 0.0}

\def\b0{\boldsymbol{0}}

\def\b0{\boldsymbol{0}}

\begin{document}
\author{Selim Gatti\footnote{RiskLab, Department of Mathematics, ETH Zurich,
selim.gatti@math.ethz.ch} \and 
  Mario V.~W\"uthrich\footnote{RiskLab, Department of Mathematics, ETH Zurich,
mario.wuethrich@math.ethz.ch}}

\date{Version of March 25, 2024}
\title{Modeling lower-truncated and right-censored insurance claims
with an extension of the MBBEFD class}
\maketitle

\begin{abstract}
\noindent  
In general insurance, claims are often lower-truncated and right-censored because 
insurance contracts may involve deductibles and maximal covers. Most classical
statistical models are not (directly) suited to model lower-truncated and right-censored claims.
A surprisingly flexible family of distributions that can cope with
lower-truncated and right-censored claims
is the class of MBBEFD distributions that originally has been introduced by Bernegger (1997) for reinsurance pricing, but which has not gained much attention outside the reinsurance literature.
Interestingly, in general insurance, we mainly rely on unimodal skewed densities, whereas the reinsurance literature typically proposes monotonically
decreasing densities within the MBBEFD class. We show that this class contains both types of densities, and we extend it
to a bigger family of distribution functions suitable for modeling lower-truncated and right-censored claims. In addition, we discuss how changes in the deductible or the maximal cover affect the chosen distributions.

~

\end{abstract}

\section{Introduction}
\begin{figure}[htb!]
\begin{center}
\begin{minipage}[t]{0.45\textwidth}
\begin{center}
\includegraphics[width=\textwidth]{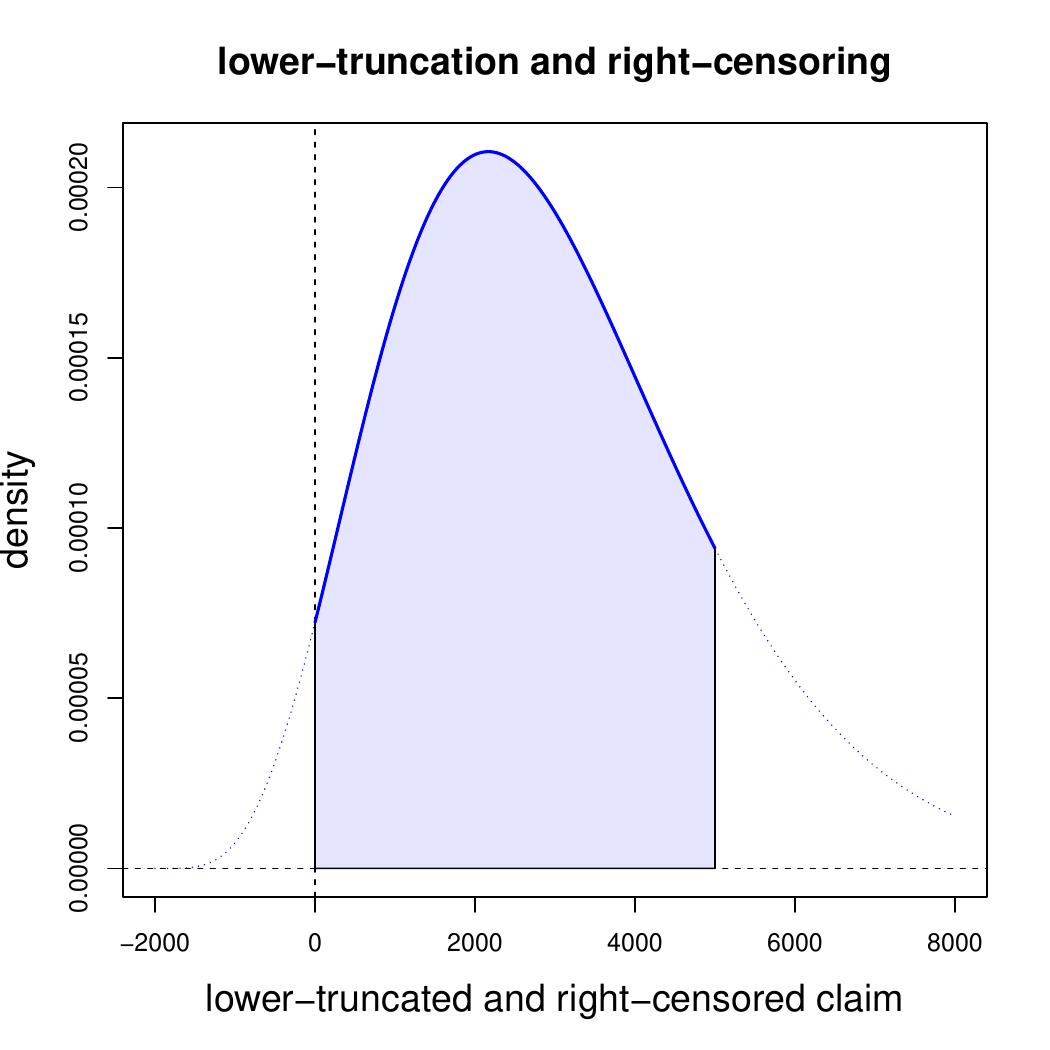}
\end{center}
\end{minipage}
\end{center}
\vspace{-.7cm}
\caption{Lower-truncated and right-censored claim
with  $d=2000$ and $M=5000$.}
\label{lower-truncation and right-censoring plot}
\end{figure}

Insurance contracts in general insurance often involve deductibles $d>0$
and maximal covers $M>0$. Deductibles are introduced to reduce the
number of small claims which mainly cause administrative expenses but which are not essential
in risk mitigation.
Maximal covers are introduced to control the maximal loss of 
an insurer. A maximal cover may, e.g., refer to the property
value insured (after subtracting the deductible), or to
the maximal insurance coverage warranted to a liability claim. Denote by $X$ 
the total financial loss. The insurance claim $Y$ after subtracting the deductible 
$d>0$ and with a maximal cover of size $M>0$ is given by
\begin{equation}
\label{claim transform}
 Y = \min\left\{(X-d)_+,\, M \right\} \, | \, X > d.
\end{equation}
We say, this financial loss is {\it lower-truncated} at $d>0$ and {\it right-censored} at $M>0$ (after subtracting the deductible).
Statistical modeling of lower-truncated and right-censored 
claims is a notoriously difficult problem. Most statistical models have an
unbounded support, e.g., the supports of the gamma and the log-normal distributions
are the entire positive real line $\R_+$. In many cases, this implies that fitting
a statistical model to
lower-truncated and right-censored data is not a problem that is easily analytically tractable.
We give an example.
We start from a classical statistical model
such as the gamma distribution for the total financial loss $X \sim F_\Gamma$, 
where $F_\Gamma$ denotes the gamma distribution
with corresponding gamma density $f_\Gamma$ on $\R_{+}$.
Lower-truncation and right-censoring introduces two difficulties
which are illustrated in Figure~\ref{lower-truncation and right-censoring plot}.
First, the lower-truncation $(X-d)_+\, | \, X > d$
of the total financial loss $X$ implies, in general, that the
density of the lower-truncated claim is positive in 0, see
Figure \ref{lower-truncation and right-censoring plot}. In the above mentioned
gamma case, this means that the lower-truncation with $d>0$ leads to a new density given by
\begin{equation*}
  y\ge 0 ~\mapsto ~
\frac{f_\Gamma(d+y)}{\int_d^\infty f_\Gamma(z)\,dz}=
\frac{f_\Gamma(d+y)}{1-F_\Gamma(d)}~>~0.
\end{equation*}
Second, right-censoring at $M>0$ of this lower-truncated claim
leads to a point mass in $M$, resulting in the density $f$ of the lower-truncated
and right-censored claim $Y$
\begin{equation}\label{lower-truncated and right-censored gamma density}
  y\ge 0 ~\mapsto ~f(y)=
\frac{f_\Gamma(d+y)}{1-F_\Gamma(d)}
\,\mathds{1}_{\{y < M\}}
+ \frac{1-F_\Gamma(d+M)}{{1-F_\Gamma(d)}}\,\mathds{1}_{\{y = M\}},
\end{equation}
where \eqref{lower-truncated and right-censored gamma density} is a density 
w.r.t.~the $\sigma$-finite measure 
being the Lebesgue measure on $(0,M)$ and having a point mass in $M$;
this point mass is not illustrated in Figure 
\ref{lower-truncation and right-censoring plot}, but only the absolutely
continuous part on $(0,M)$; the point mass in $M$ equals one minus the
volume of the blue area in Figure \ref{lower-truncation and right-censoring plot}.

More generally, for maximum likelihood estimation (MLE) based
on lower-truncated and right-censored claims $Y \in (0,M]$, $\p$-a.s.,
we consider the log-likelihood function of an unknown parameter $\theta$ given by
\begin{equation}\label{log-likelihood right-censoring}
\theta ~\mapsto~ \ell_Y(\theta) = \log \left(f_\theta(Y)\right)\mathds{1}_{\{Y < M\}}
+ \log \left(1-F_\theta(M_-)\right)\mathds{1}_{\{Y = M\}},
\end{equation}
assuming that the response variable $Y$ is absolutely continuous on
$(0,M)$ with density $f_\theta(y)$, having a point mass 
$1-F_\theta(M_-)=1-\lim_{y\uparrow M}F_\theta(y)$ in $M$, and with (unknown) model parameter~$\theta$. 
Fitting such a model with MLE can be difficult because we need an analytically tractable
form for both the density $f_\theta(\cdot)$ and its distribution function
$F_\theta(\cdot)$, see \eqref{log-likelihood right-censoring}. This is not the case, e.g., in the 
lower-truncated and right-censored gamma model given
in \eqref{lower-truncated and right-censored gamma density}.
Therefore, in such cases, one either needs to rely
on numerical integration of the density (which can be computationally demanding, e.g., when performing a regression with fixed covariates) or one uses a version of the
Expectation-Maximization (EM) algorithm by interpreting the lower-truncation
and right-censoring
as a missing information problem; we refer to 
Verbelen et al.~\cite{Verbelen}, Fung et al.~\cite{Fung2}
and Sections 6.4.2 and 6.4.3 in W\"uthrich--Merz \cite{WM2023}.
However, also this EM algorithm approach
has its drawbacks as it requires tractability of conditional tail expectations
and reasonable dispersion estimates in multi-dimensional parameter settings. These two
side constraints lead to further restrictions on the class of solvable models, e.g., 
these problems can only be solved for a very small number of models within
the class of Tweedie's models \cite{Tweedie}, namely, for the Tweedie's models stated
in Theorem 3 of Bl{\ae}sild--Jensen \cite{Blaesild}; we also
refer to Landsman--Valdez \cite{LandsmanValdez}. In a series of papers,
Poudyal \cite{Poudyal1, Poudyal2} and Poudyal--Brazauskas \cite{Poudyal3}
consider trimmed and/or winsorized methods of moments estimators
for truncated and/or censored data; in statistics, truncation is 
also called trimming and censoring winsorizing. In these papers, trimming and winsorizing is also shown to be a useful method of robustifying
moment estimation under extreme claims. 

We take a different approach in this paper to solve the fitting
problem of lower-truncated and right-censored data.
In reinsurance, often so-called MBBEFD {\it exposure curves} are used for exposure rating.
Those exposure curves have been introduced by Bernegger \cite{Bernegger}, and the acronym MBBEFD indicates that this class includes
the Maxwell--Boltzmann (MB), the Bose--Einstein (BE) and the Fermi--Dirac (FD) distributions; these are well known
distributions in statistical mechanics. These MBBEFD 
exposure curves
are based on the assumption that there is a maximal cover $M$, and they
directly describe right-censored claims up to this maximal cover. 
Differentiating twice these MBBEFD exposure curves provides us with
densities being absolutely continuous on the interval $(0,M)$ and having
a point mass in $M$; we refer to formula (3.7) in Bernegger \cite{Bernegger}.

The goal of this paper is first to study the properties of these MBBEFD densities and to extend it to a bigger class of models that will be called the \textit{Bernegger class}.
Our contribution is to show that the Bernegger class is a rich family of distributions including
monotonically decreasing densities, unimodal densities and monotonically
increasing densities, and our extension provides new families of lower-truncated and right-censored
random variables that allow for skewness in the absolutely continuous part of the distribution. This is of particular interest because unimodal skewed densities are suited
for modeling lower-truncated and right-censored claims in general insurance since their empirical density
roughly looks like the one given in Figure \ref{lower-truncation and right-censoring plot}. In particular, the distributions of lower-truncated and right-censored exponential and logistic random variables belong to the Bernegger class.

Surprisingly, the class of MBBEFD densities of Bernegger \cite{Bernegger}
has only entered the reinsurance literature; see, e.g., 
Parodi--Watson \cite{ParodiWatson}, Abramson \cite{Abramson},
Riegel \cite{Riegel}, Chapter 21 of Parodi \cite{Parodi}, and the
{\sf R} \cite{R Core Team} package {\tt mbbefd} of Dutang et al.~\cite{DutangGesmannSpedicato, SpedicatoDutang}.
Popular examples in reinsurance pricing are the so-called Swiss Re and Lloyd's exposure
curves that are special cases of MBBEFD exposure curves; see Bernegger \cite{Bernegger}. However, in this
reinsurance pricing literature, one mainly focuses on exposure curves
and not on the resulting densities nor on their properties. We show that the most popular choices
from reinsurance lead to monotonically decreasing densities, whereas we
are mainly interested into the unimodal case, as this is the common situation
in general insurance pricing. By fitting a real dataset consisting of private property insurance claims, we introduce a couple of explicit models belonging to the Bernegger class that allow for unimodal and skewed densities.

Finally, we consider the situation where the insurer is interested in understanding how a change in the deductible or the maximal cover affects the expected claim size. For this, we first emphasize that, typically, the insurer only observes the lower-truncated and right-censored claim $Y$ given in \eqref{claim transform}. That is, for statistical modeling, neither are the claims below the deductible $d$ known, nor are the exact claim sizes above the maximal cover $M$ known. Thus, we can only fit a lower-truncated and right-censored density to observations, e.g., of type \eqref{lower-truncated and right-censored gamma density}. In general, this does not allow us to extrapolate below the lower-truncation point and above the right-censoring point since there are infinitely many candidates for extrapolation. Under some assumptions on the original density and its support, we can at best smoothly extrapolate, e.g., as the dotted lines in Figure \ref{lower-truncation and right-censoring plot} suggest, but the true model could also look completely different, as $Y$ does not reveal any information about the claims being outside of its observed support, except for the proportion of claims exceeding the maximal cover. Therefore, we can only perform the opposite operation of either increasing the deductible or decreasing the maximal cover, and we will show that the Bernegger class is closed under these transformations.

\medskip

{\bf Organization.}
This manuscript is organized as follows. In the next section, we state the necessary properties that any exposure curve has to fulfill in order to describe distribution functions allowing to model lower-truncated and right-censored claims. In Section \ref{sec 3}, we start from the MBBEFD class of distributions of Bernegger \cite{Bernegger} by stating some of its properties, and then we extend it to a richer family of distributions, the Bernegger class. In Section \ref{logarithmic section}, we introduce a subclass of the Bernegger class that incorporates the logistic distribution as well as the MBBEFD class of distributions, whereas in Section \ref{exponential section}, we treat another subclass of distributions that includes the lower-truncated and right-censored exponential distribution. In Section \ref{examples}, we use a real dataset of lower-truncated and right-censored claims in order to compare the performance of the gamma and the log-normal model to five examples belonging to the Bernegger class,  fitting all these models using maximum likelihood estimation (MLE). Finally, in Section \ref{Extrapolation}, we consider the influence of a change in the deductible or the maximal cover on the observed claims distribution. The last section concludes this work. All mathematical proofs and parameters of the fitted models are provided in the appendix.

\section{Exposure curves and their resulting densities}

\subsection{From Exposure curves to distributions}
\label{Exposure curves and distributions}
In reinsurance claims modeling, one often works with exposure curves instead of distribution
functions. Assume we have a positively supported response variable $Y\sim F_Y$,
and assume that the {\it maximal possible loss} (MPL) is given by $M>0$, i.e.,
$0<Y\le M$, $\p$-a.s. We define the normalized loss $Z=Y/M$. 
Denote the distribution function of the normalized loss $Z$ by $F_Z$, being supported
in $(0,1]$. The {\it exposure curve} of a normalized loss $Z \sim F_Z$ is defined by
\begin{equation*}
z~\mapsto~  G(z) = \frac{\int_0^z 1-F_Z(s)\, ds}{\int_0^1 1-F_Z(s)\, ds}=
  \frac{\int_0^z 1-F_Z(s)\, ds}{\E[Z]},
  \end{equation*}
for $z \in [0,1]$; see Bernegger \cite{Bernegger}. This exposure curve
$G:[0,1]\to [0,1]$ is non-decreasing, concave, and satisfies the property $G'(0) > 0$, as well as the normalizations $G(0)=0$ and $G(1)=1$; for examples
see Figure \ref{Swiss Re curves} (lhs), below.
In the last integral, a change of variable $s\in [0,1] \mapsto t=s M \in [0,M]$ gives us
\begin{equation*}
  G(z) =  \frac{\int_0^{zM} 1-F_Z(t/M)\, dt}{M\E[Z]}
  = \frac{\int_0^{zM} 1-F_Y(t)\, dt}{\E[Y]} =G_Y(zM),
  \end{equation*}  
  the latter being the exposure curve of $Y$ on $[0,M]$. Thus, we can equally work with the responses $Y$ and $Z$, but the
  normalized losses $Z$ will have the advantage that they live on the common unit interval $[0,1]$. Now, let us take the opposite view and characterize the distribution function of a random variable $Z \sim F_Z$ obtained from a function $G:[0,1] \rightarrow \mathbb{R}$ satisfying the same properties of an exposure curve. Under the assumptions of the next theorem, this distribution $F_Z$ leads to an absolutely continuous density on $[0,1)$ and a point mass in 1.

\begin{theo}
    \label{Exposure curves properties}
    Let $G:[0,1] \rightarrow \mathbb{R}$ be a non-decreasing, concave, and twice continuously differentiable function with $G(0) = 0, \, G(1) = 1, \, G'(0) > 0$. The function $F_Z: [0,1] \rightarrow \mathbb{R}$ defined by
    \begin{equation}
        \label{from exposure to dist}
        F_Z(z) = \left(1- \frac{G'(z)}{G'(0)}\right) \mathds{1}_{\{z < 1\}} + \mathds{1}_{\{z = 1\}}
    \end{equation}
    is a distribution function on $[0,1]$. Furthermore, this distribution has as density
    \begin{equation}
        \label{density}
        f_Z(z) = -\frac{G''(z)}{G'(0)},
    \end{equation}
    for $z \in [0,1)$, and a point mass in 1 given by
    \begin{equation}
        \label{point mass}
        p = \frac{G'(1)}{G'(0)}.
    \end{equation}
    Finally, the mean of $Z \sim F_Z$ is equal to $\E[Z] = 1 / G'(0)$.
\end{theo}

The proofs of all statements are given in the appendix. Due to this last result, functions $G$ satisfying the assumptions of Theorem \ref{Exposure curves properties} will be called \textit{exposure curves}. 
\begin{defi}
    An exposure curve is a function $G:[0,1] \rightarrow \mathbb{R}$, which is non-decreasing, concave, and twice continuously differentiable with $G(0) = 0, \, G(1) = 1, \, G'(0) > 0$.
\end{defi}

As seen previously, if we start from any such function $G$, we can derive a distribution whose density is absolutely continuous and of closed form on $[0,1)$, with a point mass in 1 and a mean that are of closed form too, i.e., we have a class of models that has fully tractable mean, density and point mass, which is suitable to model right-censored claims. Moreover, if $G''(0)<0$, which implies $f_Z(0) > 0$, it includes lower-truncation in the sense that the density of a lower-truncated random variable $Z$ is positive in zero, see Figure \ref{lower-truncation and right-censoring plot}. The next result shows that a linear combination of exposure curves allows us to define a mixture of their respective associated distribution functions.

\begin{lemma}
    \label{Mixture distributions}
    Let $(\alpha_i)_{i=1}^n$ be non-negative weights adding up to 1 and let $(G_i)_{i=1}^n$ be exposure curves leading to distributions functions $(F_i)_{i=1}^n$, densities $(f_i)_{i=1}^n$, and point masses in 1 equal to $(p_i)_{i=1}^n$, as in Theorem \ref{from exposure to dist}. The convex combination
    \begin{equation}
        \label{mixture exposure function}
        G(z) = \sum_{i=1}^n \alpha_i G_i(z),
    \end{equation}
    for $z \in [0,1]$, is again an exposure curve allowing to define the distribution function of a random variable $Z \sim F_Z$ given by
    \begin{equation*}
        F_Z(z) = \sum_{i=1}^n w_i F_i(z),
    \end{equation*}
    for $z \in [0,1]$, and where 
    $w_i = \alpha_i G_i'(0)/\sum_{j=1}^n \alpha_j G_j'(0)$ are non-negative weights summing up to~1. In particular, the density of $Z$ on $[0,1)$ is given by
    \begin{equation*}
        f_Z(z) = \sum_{i=1}^n w_i f_i(z),
    \end{equation*}
    and the point mass in 1 is equal to
    \begin{equation*}
        p = \sum_{i=1}^n w_i p_i.
    \end{equation*}
    Finally, the mean of $Z \sim F_Z$ is given by
    \begin{equation*}
        \E[Z] = \sum_{i=1}^n w_i \, \frac{1}{G_i'(0)}.
    \end{equation*}
\end{lemma}

\subsection{Flexibility of the point mass in the right-censoring point}

\label{Flexibility of the point mass in the right-censoring point}

The point mass $p$ in 1 is automatically determined by formula \eqref{point mass}. Often, one may require more modeling flexibility
in the choice of this point mass, while still retaining the tractability of the density and the mean as it was shown in Theorem \ref{Exposure curves properties}.  A simple way to do so connects to so-called one-inflated distributions; we refer to Dutang et al.~\cite{DutangGesmannSpedicato}. In the case of 
Theorem \ref{Exposure curves properties}, this can easily be achieved. The next corollary shows, how we can obtain them by looking at the conditional density of the random variable $Z_0 \stackrel{\rm (d)}{=} Z|_{\{Z<1\}}$, which corresponds to a lower- and upper-truncated random variable.

\begin{cor} \label{General f0}
  Let $G:[0,1] \rightarrow \mathbb{R}$ be an exposure curve, we receive
  an absolutely continuous density on $[0,1)$ 
\begin{equation*}
    \begin{split}
        f_0(z)=  -\frac{1}{1-p} \frac{G''(z)}{G'(0)} = \frac{G''(z)}{G'(1) - G'(0)} ~\geq~0, 
    \end{split}
\end{equation*}
This density $f_0$ integrates to 1 and provides the mean for the random variable $Z_0\sim f_0$
  \begin{equation*}
    \E[Z_0] = \frac{1}{1-p} (\E[Z] - p) = \frac{1-G'(1)}{G'(0)-G'(1)}.
    \end{equation*}
\end{cor}

We can now add a point mass $q\in (0,1)$ in 1 to this density. This adds one more parameter to the model, giving us a mixture distribution between an absolutely continuous part $f_0$ on $[0,1)$ and a point mass in 1. We have the following corollary.

\begin{cor} \label{General f_q}
  Let $G:[0,1] \rightarrow \mathbb{R}$ be an exposure curve, the
  random variable $Z$ that has an absolutely continuous density on $[0,1)$ given
  by 
\begin{equation*}
  f_q(z)= \left(1-q\right) \frac{G''(z)}{G'(1) - G'(0)} ~\geq~0,
\end{equation*}
with fixed point mass $q\in (0,1)$ in 1 has expected value
  \begin{equation*}
    \E[Z] = \left(1-q\right)\frac{1-G'(1)}{G'(0)-G'(1)}+ q.
    \end{equation*}
\end{cor}

\medskip
Note that the transformation achieved in Corollary \ref{General f_q} can also be obtained using Lemma \ref{Mixture distributions}. Indeed, let us consider the exposure curve
\begin{equation*}
    \tilde{G}(z) = w \, G(z) + (1-w) z,
\end{equation*}
for $z \in [0,1]$, $w \in [0,1]$, and where $G$ is an exposure curve. Using Theorem \ref{Exposure curves properties}, if we denote by $F$ the distribution function obtained from the exposure curve $G$, by $f$ the absolutely continuous density on $[0,1)$ and by $p$ the point mass in~1, we can characterize the distribution function $\tilde{F}$, the absolutely continuous density on $[0,1)$ $\tilde{f}$, and the point mass $\tilde{p}$ obtained from the exposure curve $\tilde{G}$, respectively, using
\begin{equation*}
    \tilde{F}(z) = w F(z) + (1-w) \mathds{1}_{\{z=1\}},
\end{equation*}
for $z \in [0,1]$,
\begin{equation*}
    \quad \tilde{f}(z) = w f(z) = - w \frac{G''(z)}{G'(0)},
\end{equation*}
for $z \in [0,1)$, and
\begin{equation*}
    \tilde{p} = w p + (1-w).
\end{equation*}
\begin{rem}
    \textnormal{Note that similarly to Corollary \ref{General f_q}, one could also add a point mass in $0$, thus modeling lower- and right-censored insurance claims with tractable densities, means and point masses.}
\end{rem}
In what follows, the goal will be to introduce examples of exposure curves that are useful to model lower-truncated and right-censored insurance losses. For this, we start by studying the explicit family of exposure curves introduced by Bernegger \cite{Bernegger}.

\section{The Bernegger class of distributions}
\label{sec 3}

\subsection{The class of MBBEFD exposure curves and densities}
\label{The class of MBBEFD exposure curves}
The MBBEFD class of Bernegger \cite{Bernegger} selects an
explicit family of exposure curves.
This family is characterized through two parameters $g\ge 1$ and $b\ge 0$, and it is
given as follows for $z \in [0,1]$,
\begin{equation}\label{MBBEFD exposure curve definition}
  G_{b,g}(z) = \left\{
    \begin{array}{ll}
      z & \text{ for $g=1$ or $b=0$,}\\
 \frac{\log(1+(g-1)z)}{\log(g)} & \text{ for $g>1$ and $b=1$,}\\  
 \frac{1-b^z}{1-b} & \text{ for $g>1$ and $bg=1$,}\\  
      \frac{\log\left(\frac{(g-1)b+(1-bg)b^z)}{(1-b)}\right)}{\log(bg)} & \text{ for $g>1$, $b>0$, $b\neq 1$ and $bg\neq 1$.}
    \end{array}
  \right.
\end{equation}
The three cases $bg=1$, $bg>1$ and $bg<1$ give the MB (Maxwell-Boltzmann), the BE (Bose-Einstein) and the 
FD (Fermi-Dirac) distributions, respectively.
Using \eqref{from exposure to dist}, we can calculate 
 the distributions for $z \in [0,1)$, see (3.6) in Bernegger \cite{Bernegger},
\begin{equation} \label{distribtuion MBBEFD}
  F_{b,g}(z) = \left\{
    \begin{array}{ll}
      0 & \text{ for $g=1$ or $b=0$,}\\
 1-\left(1+(g-1)z\right)^{-1} & \text{ for $g>1$ and $b=1$,}\\  
 1-b^z & \text{ for $g>1$ and $bg=1$,}\\  
 1-\frac{1-b}{(g-1)b^{1-z}+(1-bg)}      & \text{ for $g>1$, $b>0$, $b\neq 1$ and $bg\neq 1$.}
   \end{array}
  \right. 
\end{equation}
We observe that the first case is not of interest
because it gives a point mass of 1 to $z=1$. For this reason, we skip this case in the sequel.
On $z<1$, we can calculate the second derivatives of these exposure curves $G_{b,g}$. This gives us the
densities for $z\in[0,1)$, see (3.7) in Bernegger \cite{Bernegger},
\begin{equation}\label{density MBBEFD}
  f_{b,g}(z) = \left\{
    \begin{array}{ll}
 (g-1)\left(1+(g-1)z\right)^{-2} & \text{ for $g>1$ and $b=1$,}\\  
 -\log(b)b^z & \text{ for $g>1$ and $bg=1$,}\\  
 \frac{(g-1)(b-1)\log(b)b^{1-z}}{\left((g-1)b^{1-z}+(1-bg)\right)^2}      & \text{ for $g>1$, $b>0$, $b\neq 1$ and $bg\neq 1$,}
    \end{array}
  \right.
\end{equation}
and we have a point mass in $z=1$ given by
\begin{equation*}
  p=\frac{1}{g} ~\in~ (0,1).
\end{equation*}

Thus, we have an absolutely continuous distribution
on $[0,1)$, with a point mass $p=1/g$ in 1,  and, 
e.g., in the last case of \eqref{density MBBEFD}, we have a strictly positive density in 0
\begin{equation*}
f_{b,g}(0) = 
 \frac{(g-1)\log(b)b}{b-1}~>~0.
\end{equation*}
Such a density may therefore come from a lower-truncated claim.
Finally, the mean is given by
\begin{equation*}
  \E_{b,g}[Z] = \left\{
    \begin{array}{ll}
 \frac{\log(g)}{g-1} & \text{ for $g>1$ and $b=1$,}\\  
 \frac{b-1}{\log(b)} & \text{ for $g>1$ and $bg=1$,}\\  
      \frac{b-1}{\log(b)}\frac{\log(bg)}{bg-1} & \text{ for $g>1$, $b>0$, $b\neq 1$ and $bg\neq 1$.}
    \end{array}
  \right.
\end{equation*}

We give an example of such an exposure curve that is typically used
for exposure rating in reinsurance.

\begin{figure}[htb!]
\begin{center}
\begin{minipage}[t]{0.45\textwidth}
\begin{center}
\includegraphics[width=\textwidth]{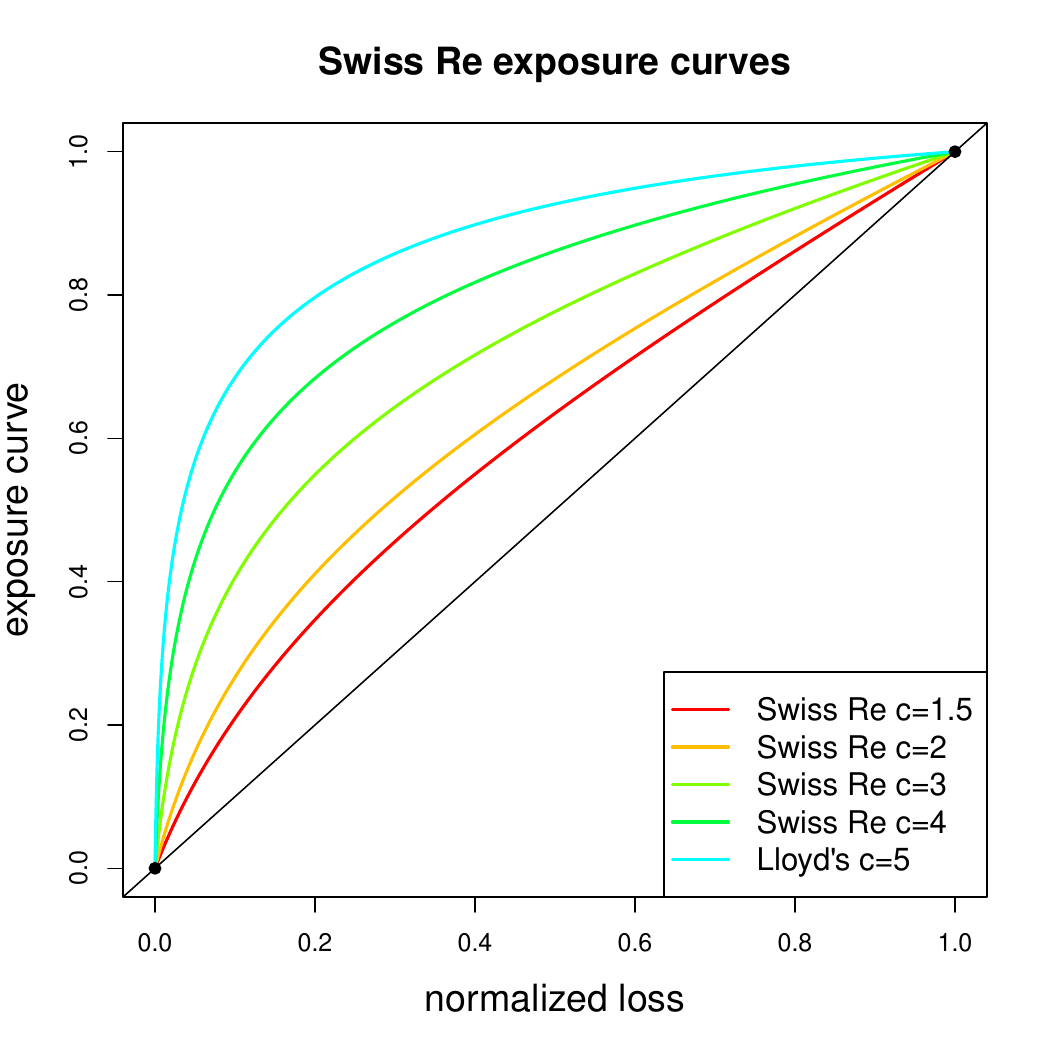}
\end{center}
\end{minipage}
\begin{minipage}[t]{0.45\textwidth}
\begin{center}
\includegraphics[width=\textwidth]{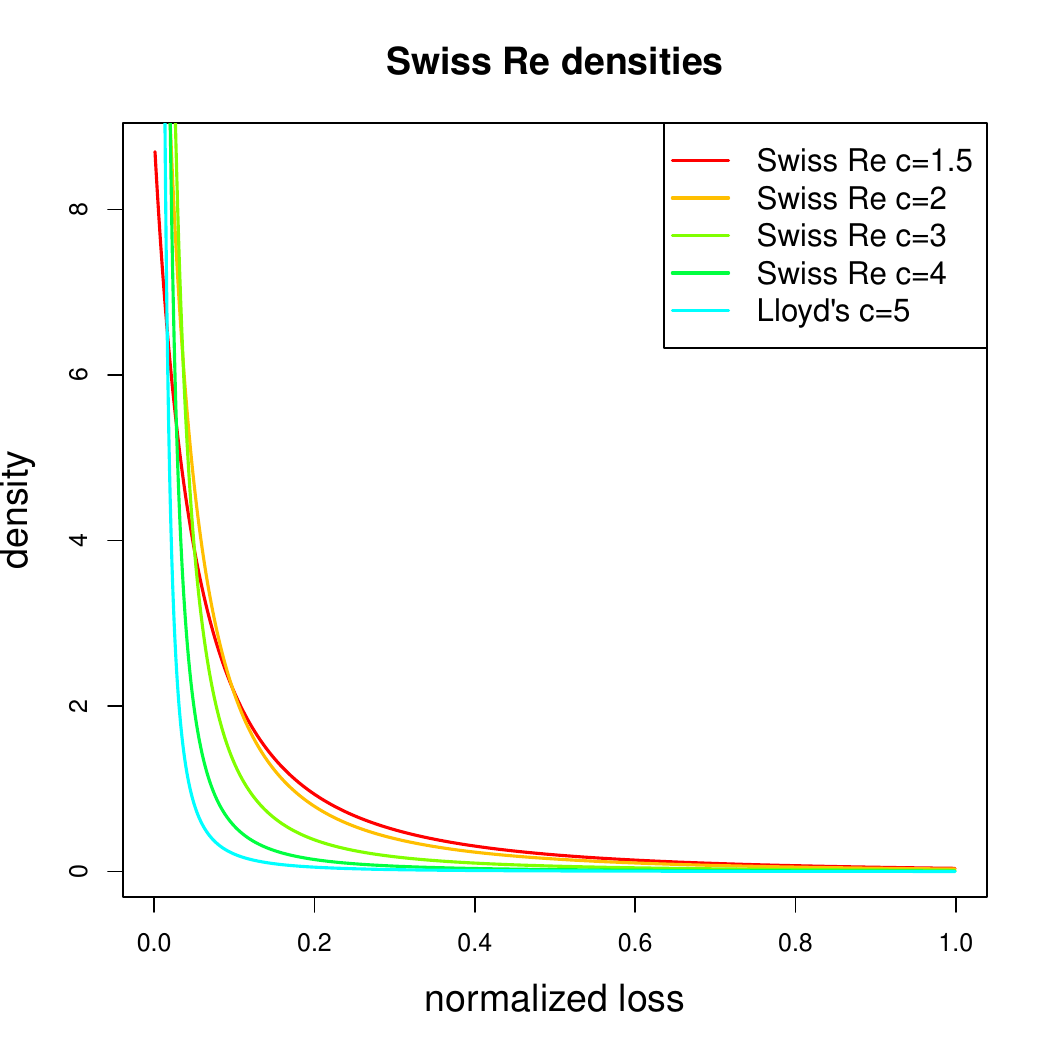}
\end{center}
\end{minipage}
\end{center}
\vspace{-.7cm}
\caption{Swiss Re and Lloyd's exposure curves (lhs) and the resulting
densities (rhs).}
\label{Swiss Re curves}
\end{figure}

\begin{example}[Swiss Re and Lloyd's exposure curves]
\normalfont \label{Swiss Re example}
Bernegger \cite{Bernegger} provides an explicit parametrization for the
MBBEFD class which can be used for reinsurance exposure rating in case of
scarce data. Namely, both parameters $b$ and $g$ are parametrized as
a function of a single parameter $c>0$ as follows
\begin{equation} \label{Swiss Re}
b=b(c)=\exp\{3.1 - 0.15 (1+c)c\} \qquad \text{ and }
\qquad g=g(c)=\exp\{(0.78 + 0.12c)c\}.
\end{equation}
For $c=1.5, 2, 3, 4$, one obtains the Swiss Re exposure curves, and for $c=5$,
the Lloyd's exposure curve, these are illustrated in Figure \ref{Swiss Re curves} (lhs).
The right-hand side of this figure shows the resulting densities $f_{b(c),g(c)}(\cdot)$, and 
we remark that all of the considered MBBEFD densities 
 are monotonically decreasing on $[0,1]$.
\EndExample
\end{example}

\subsection{Properties of the MBBEFD class}

Example \ref{Swiss Re example} has shown that the exposure
curves \eqref{Swiss Re} proposed by Bernegger \cite{Bernegger} for reinsurance exposure
rating lead to monotonically decreasing densities $f_{b,g}$, see Figure \ref{Swiss Re curves} (rhs). For general
insurance pricing, we are rather interested into unimodal
densities similar to Figure \ref{lower-truncation and right-censoring plot}, because
this more commonly reflects the properties of general insurance claims data.

\begin{prop}\label{analysis of modes}
The density $f_{b,g}$ for $g>1$ and $b>0$ given in \eqref{density MBBEFD} has the following properties:
\begin{itemize}
\item Case $bg<1$. The density $f_{b,g}$ is 
\begin{itemize}
\item monotonically decreasing on $[0,1)$ for $(1-bg)/(g-1)\le b$;
\item unimodal on $[0,1)$ for $b<(1-bg)/(g-1)<1$,
with a maximum in
\begin{equation}\label{maximum z star}
z^*=1-\frac{\log\left((1-bg)/(g-1)\right)}{\log(b)}~\in~(0,1);
 \end{equation}
\item monotonically increasing on $[0,1)$ for $(1-bg)/(g-1)\ge 1$.
 \end{itemize}
\item Case $bg\ge 1$. The density $f_{b,g}$ is monotonically decreasing on $[0,1)$.
\end{itemize}
\end{prop}

This proposition shows that in practical applications
in general insurance, the FD distributions
(with $bg<1$) are the most interesting ones, as they can be unimodal, or either
monotonically decreasing or increasing. This excludes the
exposure curves of Example \ref{Swiss Re example}, as these
Swiss Re and Lloyd's exposure curves provide us with $bg>1$.
Next, we show that the MBBEFD distribution $F_{b,g}$ for $bg < 1$
can be derived from the logistic function (distribution)
\begin{equation*}
  \psi(t) = \frac{e^t}{e^t+1} ~\in~(0,1),
\end{equation*}
for $t\in \R$. The logistic function has first derivative (logistic density)
\begin{equation*}
  \psi'(t)= \frac{e^t}{(e^t+1)^2}=\psi(t)\left(1-\psi(t)\right).
\end{equation*}
This derivative is symmetric around zero, which leads to the following result.

\begin{prop}
\label{prop bell shaped}
Let $b > 0$ and $g > 1$. For $bg < 1$, the MBBEFD density has the 
functional form, for $z\in [0,1)$,
\begin{equation*}
  f_{b,g}(z) = (a+1)\log(1/b)\, \psi'\left(z \log(1/b)+\log(a)\right),
\end{equation*}
where we set
\begin{equation}\label{reparametrization 1}
  a=\frac{b(g-1)}{1-bg} \qquad \text{ and } \qquad
g=\frac{a+b}{(a+1)b},
\end{equation}
respectively.
If $b<(1-bg)/(g-1)<1$, i.e. $b<a<1$, this MBBEFD density is
bell shaped around $z^*$ given in \eqref{maximum z star}.
\end{prop}

For further terminology, we call unimodal symmetric densities, bell shaped densities. Of course, this includes for example the Gaussian density but also the logistic density. We conclude that for $bg<1$, the MBBEFD density is the logistic density $\psi'$ on the interval
\begin{equation*}
  \left[\log(a),\,
    \log(a/b)\right),
\end{equation*}
scaled by a constant factor $(1+a)\log(1/b)>0$. It is symmetric around its mode $z^*$, and it decays slower than the Gaussian
density. This now shows why the MBBEFD densities are not sufficient for general insurance claims modeling, because general insurance claims are typically positively skewed, which cannot be captured by the logistic density.

\subsection{Extension of Bernegger's idea using non-bell shaped densities}
\label{Extending the idea of MBBEFD exposure curves}

We extend the class of bell-shaped MBBEFD densities to a more general class of exposure curves, which allows, in particular, for skewness in their corresponding densities. We call this extended family the Bernegger class.
In Section \ref{Exposure curves and distributions}, we have started from a generic
exposure curve $G:[0,1]\to [0,1]$ which is a non-decreasing, concave and twice continuously differentiable function with the normalizations $G(0)=0, \, G(1)=1$ and with $G'(0)>0$.
The MBBEFD exposure curve \eqref{MBBEFD exposure curve definition} can be reparametrized. Indeed,
by using \eqref{reparametrization 1},  we obtain for $z \in [0,1]$ 
\begin{equation}\label{MBBEFD exposure curve a and b}
  G_{b,a}(z) = \frac{\log(a+b^z)-\log(a+1)}{\log(a+b)-\log(a+1)},
\end{equation}
for parameters $b \geq 0$ and $a > -\min (1, b)$ chosen such that $G_{b,a}$ is an exposure curve; we refer to Section 3.1 of Bernegger \cite{Bernegger}. This structure can be used to design exposure curve forms that
do not have the bell-shape property of Proposition~\ref{prop bell shaped}.
We modify the modeling set-up \eqref{MBBEFD exposure curve a and b} as follows. Choose a function $B: [0,1] \rightarrow \R$ that satisfies 
\begin{equation}
\label{inner link}
    B(z) = h(b(z)),
\end{equation}
for some functions $h$ and $b$, in order to define an exposure curve (under further assumptions on $h$ and $b$)
\begin{equation}
\label{general G}
    z \mapsto G(z) = \frac{B(z)-B(0)}{B(1)-B(0)},
\end{equation}
which ensures that the normalization property $G(0) = 0$ and $G(1) = 1$ is satisfied.
The function $h$ will be denoted as the \textit{link function}, whereas the function $b$ will be named the \textit{inner function}, and we notice that Bernegger's original choice was $b(z) = a + b^z$ and $h(x) = \log(x)$, meaning that he used a \textit{logarithmic linked exposure curve}. We will first explore some examples using the same link function and then introduce the \textit{exponentially linked exposure curves}, which use the link function $h(x) = \exp(x)$. We call the class of distributions induced by exposure curves of the form \eqref{general G} the Bernegger class.

\section{Logarithmic linked exposure family}

\label{logarithmic section}

We start by considering logarithmic linked examples of the Bernegger class, which are obtained by choosing $h(x) = \log(x)$ in \eqref{inner link}.


\begin{prop}
    \label{prop exp curve 1}
    Choose a function $b:[0,1]\to (0,\infty)$ with $b(0) \neq b(1)$ that is twice continuously differentiable and define for $z \in [0,1]$ the function
    \begin{equation}
        \label{exposure curve 1}
        G(z) = \frac{\log (b(z))-\log(b(0))}{\log(b(1))-\log(b(0))}.
    \end{equation}
    The function $G$ is an exposure curve if and only if one of the following two holds:
    \begin{equation}
        \label{log link cond 1}
        b'(0) > 0, \, b'(z) \geq 0 \textrm{ and } b''(z)b(z) - b'(z)^2 \leq 0 \textrm{ for all } z \in [0,1],
    \end{equation}
    or 
    \begin{equation}
        \label{log link cond 2}
        b'(0) < 0, \,  b'(z) \leq 0 \textrm{ and } b''(z)b(z) - b'(z)^2 \geq 0 \textrm{ for all } z \in [0,1].
    \end{equation}
\end{prop}

Using Theorem \ref{Exposure curves properties}, one can then derive the distribution function of a random variable leading to an absolutely continuous density on $[0,1)$ and a point mass in 1.

\begin{cor} \label{cor exp curve 1}
Assume that a twice continuously differentiable function $b:[0,1] \to (0,\infty)$ with $b(0)~ \neq~b(1)$ fulfills condition \eqref{log link cond 1} or \eqref{log link cond 2}. The exposure curve $G$ defined in \eqref{exposure curve 1} provides the distribution of a random variable $Z \sim F_Z$
\begin{equation}
\label{distfct}
 F_Z(z) = \left(1-\frac{b'(z)}{b'(0)} \frac{b(0)}{b(z)} \right) \mathds{1}_{\{z < 1\}}+ \mathds{1}_{\{z = 1\}},
\end{equation}
 for $z\in[0,1]$, with density for $z\in [0,1)$
\begin{equation*}
  f_Z(z)= \frac{b(0)}{-b'(0)} \frac{b''(z) b(z) - b'(z)^2}{b(z)^2},
\end{equation*}
and with point mass in $z=1$ equal to
\begin{equation*}
  p = \frac{b'(1)}{b'(0)} \frac{b(0)}{b(1)}.
\end{equation*}
Moreover, the mean of $Z \sim F_Z$ is equal to
\begin{equation*}
  \E[Z] = \frac{b(0)}{-b'(0)}\, \log\left(\frac{b(0)}{b(1)}\right).
\end{equation*}
\end{cor}

\medskip

For general insurance pricing, we are interested into unimodal densities and we can thus derive the first derivative of $f_Z$ in order to characterize the maximum of the density
\begin{equation*}
    f'_Z(z) =   \frac{b(0)}{-b'(0)} \frac{b'''(z) b(z)^2 -3b''(z)b'(z)b(z) + 2 b'(z)^3}{b(z)^3}.
\end{equation*}
Note that this derivative only exists if the third derivative of $b$ exists. The next result shows an explicit member of the Bernegger class that belongs to the logarithmic linked exposure family.

\begin{example}[Two-parameter logistic distribution]

\normalfont \label{Logistic distribution}
    Consider a random variable $X$ following a two-parameter logistic distribution with density
\begin{equation*}
    f_{X}(z) = \frac{e^{(z-\mu)/\sigma}}{\sigma \left( 1 + e^{(z-\mu)/\sigma} \right)^2}, \quad \textrm{for} \, -\infty < z < \infty, \, -\infty < \mu < \infty, \, \sigma > 0,
\end{equation*}
and distribution
\begin{equation*}
    F_{X}(z) = \frac{e^{(z-\mu)/\sigma}}{1 + e^{(z-\mu)/\sigma}}, \quad \textrm{for} \, -\infty < z < \infty, \, -\infty < \mu < \infty, \, \sigma > 0.
\end{equation*}
Let $d \in \R$ and $M > 0$ in order to define the scaled lower-truncated and right-censored random variable
\begin{equation}
    \label{lower-truncated and right-censored logistic}
    Z = \frac{1}{M} \min \{(X-d)_{+}, M \} \, | \, X > d.  
\end{equation}
The distribution of $Z$ is given by
\begin{equation}
 \label{dist of X Z}
    F_{Z}(z) = \frac{F_X(d+zM) - F_X(d)}{1-F_X(d)}
\,\mathds{1}_{\{z \in [0,1) \}} + \mathds{1}_{\{z =1\}},
\end{equation}
which implies
\begin{equation*}
    F_{Z}(z) = \frac{e^{(d+zM-\mu)/\sigma}-e^{(d-\mu)/\sigma}}{1 + e^{(d+zM-\mu)/\sigma}}
\,\mathds{1}_{\{z \in [0,1) \}} + \mathds{1}_{\{z = 1\}}, \\
\end{equation*}
for $z \in [0,1]$. Furthermore, we have
\begin{equation*}
        \int_0^z 1- F_{Z}(s) \, ds = \frac{-\sigma}{M} \left(1+e^{(d-\mu)/\sigma}\right)\left[ \log \left( 1 +  e^{-(d+zM-\mu)/\sigma}\right) - \log \left( 1 + e^{-(d-\mu)/\sigma}\right) \right],
\end{equation*}
for $z \in [0,1]$. This implies that the exposure curve of $Z$ is given by
\begin{equation*}
    \begin{split}
        G(z) &= \frac{\int_0^z 1-F_Z(s)\, ds}{\int_0^1 1-F_Z(s)\, ds} \\
        &= \frac{\log \left(1 + e^{-(d+zM-\mu)/\sigma}\right) - \log \left( 1 + e^{-(d-\mu)/\sigma}\right)}{\log \left( 1 + e^{-(d+M-\mu)/\sigma}\right) - \log \left( 1 + e^{-(d-\mu)/\sigma}\right)},
    \end{split}
\end{equation*}
for $z \in [0,1]$, which shows that the distribution of a (scaled) lower-truncated and right-censored two-parameter logistic random variable belongs to the Bernegger class with a logarithmic link function $h(x) = \log(x)$ and inner function $b(z) = 1 + e^{-(d+zM-\mu)/\sigma}$.
\EndExample
\end{example}

\section{Exponentially linked exposure family}

\label{exponential section}

Next we introduce the exponentially linked exposure family by setting $h(x) = \exp(x)$ in \eqref{inner link}.


\begin{prop}
    \label{prop exp curve 2}
    Choose a function $b:[0,1]\to \R$ with $b(0) \neq b(1)$ that is twice continuously differentiable and define for $z \in [0,1]$ the function
    \begin{equation}
        \label{exposure curve 2}
        G(z) = \frac{e^{b(z)}-e^{b(0)}}{e^{b(1)}-e^{b(0)}}.
    \end{equation}
    The function $G$ is an exposure curve if and only if one of the following two holds:
    \begin{equation}
        \label{exp link cond 1}
        b'(0) > 0, \, b'(z) \geq 0 \textrm{ and } b''(z) + b'(z)^2 \leq 0 \textrm{ for all } z \in [0,1],
    \end{equation}
    or 
    \begin{equation}
        \label{exp link cond 2}
        b'(0) < 0, \,b'(z) \leq 0 \textrm{ and } b''(z) + b'(z)^2 \geq 0 \textrm{ for all } z \in [0,1].
    \end{equation}
\end{prop}

As for the logarithmic linked exposure curves, one can then derive, using Theorem \ref{Exposure curves properties}, a distribution function leading to an absolutely continuous density on $[0,1)$ and to a point mass in~1.

\begin{cor} \label{cor exp curve 2}
Assume that a twice continuously differentiable function $b:[0,1] \to \R$ with $b(0) \neq b(1)$ fulfills condition \eqref{exp link cond 1} or \eqref{exp link cond 2}. The exposure curve $G$ defined in \eqref{exposure curve 2} provides the distribution of a random variable $Z \sim F_Z$
\begin{equation}
\label{distfct 2}
 F_Z(z) = \left(1- \frac{e^{b(z)}}{e^{b(0)}} \frac{b'(z)}{b'(0)}\right) \mathds{1}_{\{z < 1\}} + \mathds{1}_{\{z = 1\}},
\end{equation}
 for $z\in[0,1]$, with density for $z\in [0,1)$
\begin{equation}
    \label{density 2}
  f_Z(z) = - \frac{e^{b(z)}}{e^{b(0)}}  \frac{b'(z)^2 + b''(z)}{b'(0)},
\end{equation}
and with point mass in $z=1$ equal to
\begin{equation*}
  p = \frac{e^{b(1)}}{e^{b(0)}} \frac{b'(1)}{b'(0)}.
\end{equation*}
The mean of $Z \sim F_Z$ is equal to
\begin{equation*}
  \E[Z] = \frac{e^{b(1)-b(0)} - 1}{b'(0)}.
\end{equation*}
\end{cor}

\medskip

The first derivative of the density $f_Z$ given in \eqref{density 2} allows us to characterize its extrema and is given by
\begin{equation}
    \label{derivative exp curve 2}
    f_Z'(z) = - \frac{e^{b(z)}}{e^{b(0)}}  \frac{b'(z)^3 + 3b'(z)b''(z)+b'''(z)}{b'(0)}.
\end{equation}
Note that this derivative only exists if the third derivative of $b$ exists. The next example shows that lower-truncated and right-censored exponential random variables belong to the exponentially linked exposure family.

\begin{example}[Exponential distribution]

\normalfont \label{Exponential distribution}
    Consider a total financial loss $X \sim \textrm{Exp}(\lambda)$ as well as a deductible $d > 0$ and a maximal cover $M > 0$. Then the scaled lower-truncated and right-censored insurance claim is given by
    \begin{equation*}
        Z = \frac{1}{M} \min\left\{(X-d)_+,\, M \right\} \, | \, X > d,
    \end{equation*}
    and its distribution reads as
    \begin{equation*}
        \begin{split}
            F_{Z}(z) &= \frac{F_X(d+zM) - F_X(d)}{1-F_X(d)} \,\mathds{1}_{\{z \in [0,1) \}} + \mathds{1}_{\{z = 1\}} \\
            &= \left(1 - e^{- \lambda zM} \right) \,\mathds{1}_{\{z \in [0,1) \}} + \mathds{1}_{\{z = 1\}},
        \end{split}
    \end{equation*}
    for $z \in [0,1]$. This implies that the exposure curve of the random variable $Z$ is given by
    \begin{equation*}
        G(z) = \frac{\int_0^z 1-F_Z(s)\, ds}{\int_0^1 1-F_Z(s)\, ds}= \frac{e^{- \lambda zM} - 1}{e^{- \lambda M} - 1},
    \end{equation*}
    for $z \in [0,1]$, which shows that the distribution of a (scaled) lower-truncated and right-censored exponential random variable belongs to the Bernegger class with an exponential link function~$h(x) = \exp(x)$ and a linear inner function $b(z) = - \lambda zM$, where $M$ stands for the maximal cover.
\EndExample
\end{example}

We point out that the exponentially linked exposure family is equal to the logarithmic linked exposure family as stated in the next proposition. Thus, the choice of using one family over the other is mainly motivated by having simpler forms in the inner function $b$.
\begin{prop}
\label{log family = exp family}
    The logarithmic linked exposure family and the exponentially linked exposure family coincide.
\end{prop}

\section{Real dataset example}

\label{examples}

In this section, the goal is to exploit some examples belonging to the logarithmic and the exponentially linked exposure family of the Bernegger class. These examples will be used to fit general insurance claims data using the tractability of the models described in Section \ref{Exposure curves and distributions}. In other words, MLE can directly be used since the distribution functions as well as their associated densities are of closed form.

For this, we will vary the choice of the inner function $b(z)$. In the following, $\theta \in \Theta$ will denote the set of parameters appearing in the inner function. Using a dataset for $Z$ taking values in~$(0,1]$, we will fit the models with MLE, maximizing the log-likelihood function
\begin{equation}
    \label{log lik 1}
    \theta \mapsto \ell_{Z}(\theta) = \log(f^{(\theta)}(Z)) \mathds{1}_{\{Z<1\}} + \log(p^{(\theta)}) \mathds{1}_{\{Z=1\}},
\end{equation}
where the absolutely continuous density $f^{(\theta)}$ on $[0,1)$ and the point mass $p^{(\theta)}$ are obtained as in Section \ref{Exposure curves and distributions}. We also study the model of Corollary \ref{General f_q}, which extends the previous model by adding a flexible point mass $q$ in 1. Its log-likelihood function is
\begin{equation}
    \label{log lik 2}
    \begin{split}
        (\theta, q)\mapsto \ell_{Z}(\theta, q) &= \log((1-q) f^{(\theta)}_0(Z)) \mathds{1}_{\{Z<1\}} + \log(q) \mathds{1}_{\{Z=1\}} \\
     &= \log(f^{(\theta)}_0(Z)) \mathds{1}_{\{Z<1\}} + \log(1-q) 
     \mathds{1}_{\{Z<1\}} + \log(q) \mathds{1}_{\{Z=1\}}.
    \end{split}
\end{equation}

The maximization problem in \eqref{log lik 1} will be denoted as the \textit{MLE of the standard problem}, whereas the maximization problem in \eqref{log lik 2} will be called the \textit{MLE of the extended problem}.

The dataset used in this section consists of claims observations from private property insurance. Private property usually includes deductibles to reduce the number of small claims, and hence, administrative expenses, e.g. , a sufficiently high deductible implies that not every lost umbrella gets reported to the insurance company as stolen. Secondly, private property includes maximal covers that may depend on the underlying peril. We have $n = 126\,026$ claims $Y_i$ above the deductible $d$ and we scale them by the maximal cover providing us with normalized lower-truncated and right-censored claims $Z_i$ for $i = 1, \dots, n$. We assume that these normalized claims $Z_i$ are i.i.d. and follow a distribution belonging to the Bernegger class.

\begin{figure}[htb!]
\begin{center}
\begin{minipage}[t]{0.45\textwidth}
\begin{center}
\includegraphics[width=\textwidth]{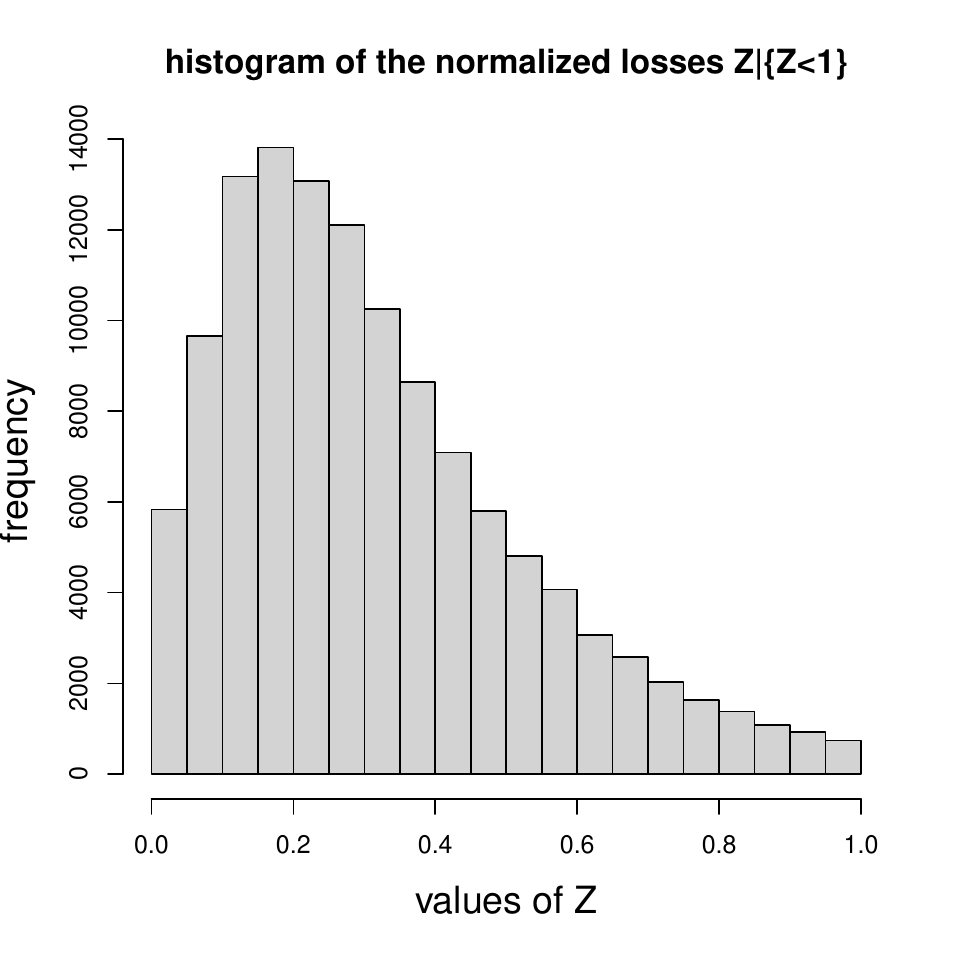}
\end{center}
\end{minipage}
\begin{minipage}[t]{0.45\textwidth}
\begin{center}
\includegraphics[width=\textwidth]{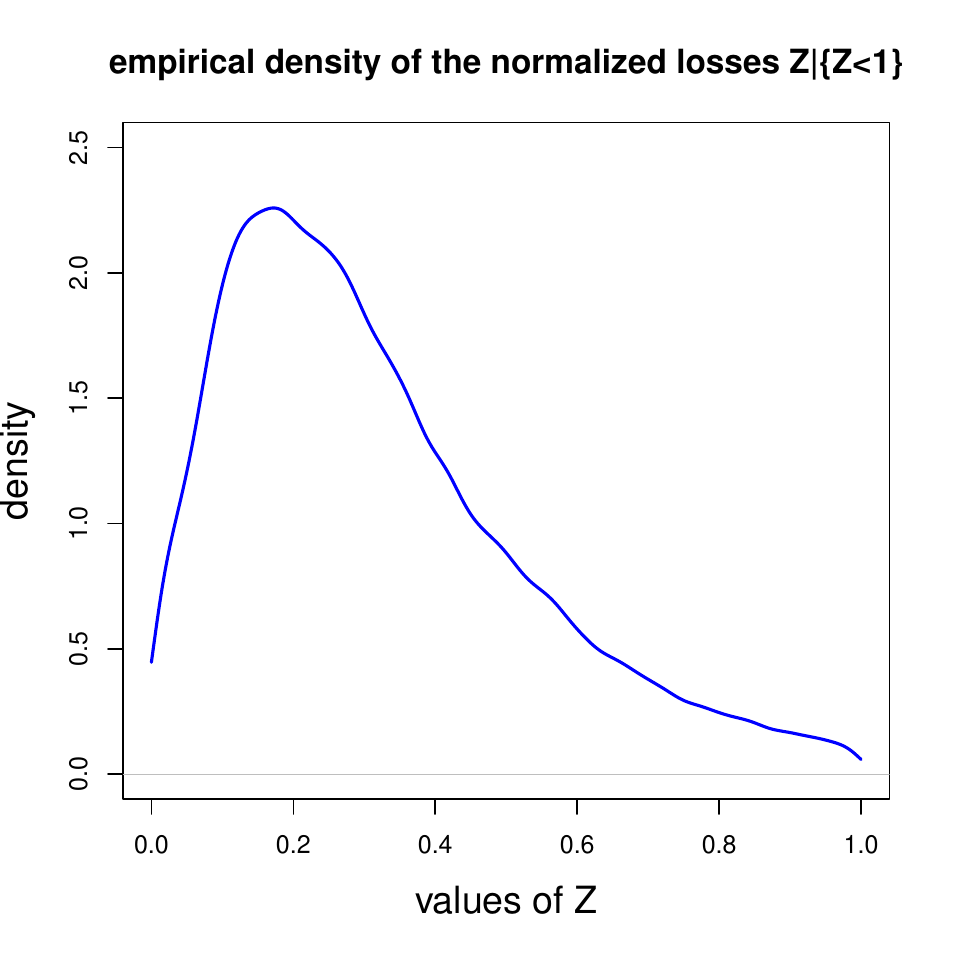}
\end{center}
\end{minipage}
\end{center}
\vspace{-.7cm}
\caption{Histogram (lhs) and empirical density (rhs) of the claims $Z_i$, only showing the claims strictly smaller than $1$, and the point mass in $1$ is $3.4 \%$.}
\label{Dataset used}
\end{figure}

\begin{table}[htb!]
    \centering
    \begin{tabular}{ l c c c c c c}
      \toprule
       & Min. & Q1 & Median & Q3 & Max. & Mean \\
      \midrule
      Normalized claims $Z_i$ & 0.00001 & 0.160 & 0.280 & 0.457 & 1 & 0.339 \\
      \bottomrule
    \end{tabular}
    \caption{Summary statistics of the dataset containing $126\,026$ normalized lower-truncated and right-censored claims $Z_i$.}
    \label{Tab:Summary statistics}
\end{table}
 Some summary statistics of the claims $Z_i$ are provided in Table \ref{Tab:Summary statistics} and the empirical (observed) density of the claims that are strictly smaller than $1$ is shown in Figure \ref{Dataset used}, we have an empirical mean of $0.339$ and the observed point mass in 1 is $0.034$, i.e., the insurance company pays the maximal cover on $3.4 \%$ of the claims.

These normalized losses $Z_i$, along with different models from the Bernegger class, will be used to solve the above MLE maximization problems in order to produce the results of this section. For this, we use the {\sf R} function {\it optim} in order to minimize the sum of the negative of the log-likelihoods evaluated at $Z_i$, after having possibly transformed our parameters in a way to ensure that they lie in a suitable open domain, this is described below. The results of the best model from the Bernegger class will then be compared to lower-truncated and right-censored log-normal and gamma models at the end of this section. The first fitted model is the classical MBBEFD model of Bernegger \cite{Bernegger}.

\subsection{The MBBEFD example}

\label{MBBEFD}

\begin{figure}[htb!]
    \centering
    \includegraphics[scale=0.7]{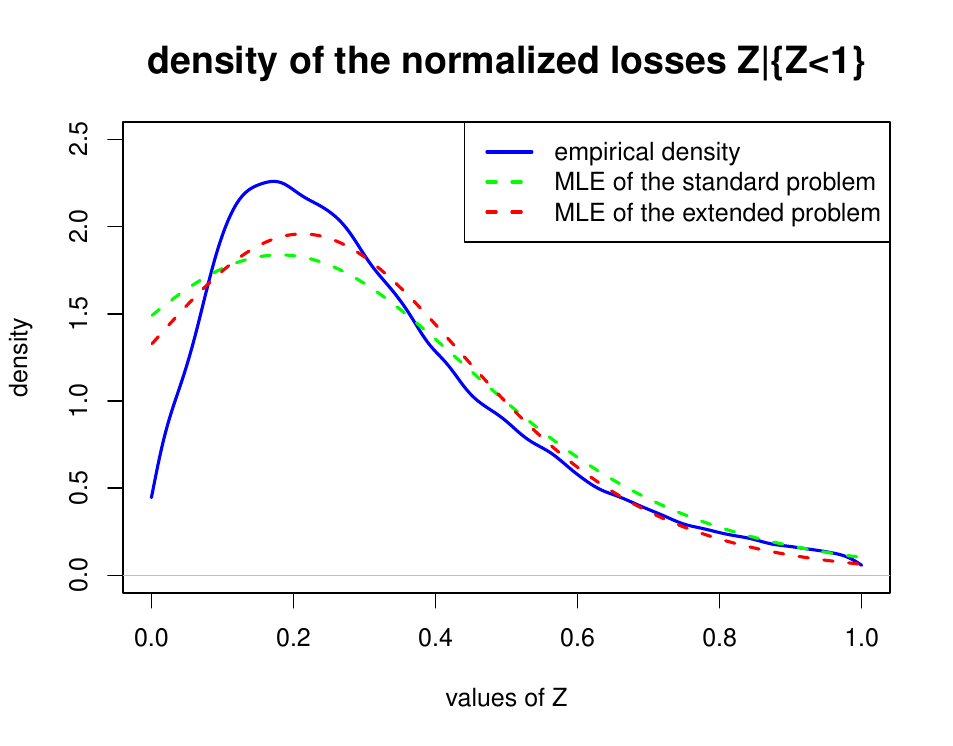}
    \caption{The MBBEFD example: densities of the random variable $Z|_{\{Z<1\}}$.}
    \label{Fig:MBBEFD}
\end{figure}

We have seen in Section \ref{Extending the idea of MBBEFD exposure curves} that the MBBEFD example belongs to the logarithmic linked exposure family and is obtained by choosing the inner function
\begin{equation*}
  b(z)= a + b^z,
\end{equation*}
for parameters $b \geq 0$ and $a > -\min (1, b)$ chosen in a way to obtain a well-defined exposure curve. Using the parametrization in \eqref{reparametrization 1}, it is possible to give the conditions under which this class of distributions leads to unimodal densities on $[0,1)$. Indeed, according to Proposition \ref{analysis of modes}, such unimodal densities are obtained if and only if
\begin{equation*}
    bg < 1
  \qquad \text{ and } \qquad
   b < \frac{1-bg}{g-1} < 1,
\end{equation*}
for parameters $g > 1$ and $b > 0$. Therefore, we use the domain
\begin{equation*}
    \Theta = \Big\{ g > 1, \max \left(0,\frac{2-g}{g} \right) < b < \frac{1}{2g-1} \Big\},
\end{equation*}
in order to find unimodal MLE solutions of the standard and extended maximization problems described in \eqref{log lik 1} and \eqref{log lik 2}. In Figure \ref{Fig:MBBEFD}, the empirical density of the data points strictly smaller than 1 is plotted in blue color. Since this corresponds to a true density integrating to one, we further show the conditional density of $Z|_{\{Z<1\}}$ of the fitted models, using the parameters obtained by solving the standard problem \eqref{log lik 1} (in green) and the extended problem \eqref{log lik 2} (in red). As for all the fitted models in this section, these parameters are provided in the appendix.. The point mass and the mean are shown in Table~\ref{Tab:MBBEFD}, as well as the log-likelihoods of the random variables $Z|_{\{Z<1\}}$ and $Z$, and the AIC scores that are computed using $\ell_Z$. We see that, as expected, the MLE solution of the extended problem gives better results, even if the densities obtained are not close to the empirical density. Proposition \ref{prop bell shaped} helps us to understand why the fit is not accurate since the empirical density is skewed to the right, whereas the MBBEFD class only allows for symmetric densities described by the derivative of the logistic function.

\begin{table}[htb!]
    \centering
    \begin{tabular}{ l c c c c c}
      \toprule
       & Point mass & Mean & $\ell_{Z|_{\{Z<1\}}}$ & $\ell_Z$ & AIC \\
      \midrule
      Empirical density (Blue) & 0.034 & 0.339 & - & - & - \\
      MLE of the standard problem (Green) & 0.020 & 0.337 &32 682 &13 475 & -26 947 \\
      MLE of the extended problem (Red) & 0.034 & 0.336 &33 257 &14 587 & -29 168 \\
      \bottomrule
    \end{tabular}
    \caption{The MBBEFD example: results.}
    \label{Tab:MBBEFD} 
\end{table}

\subsection{Power logarithmic linked exposure example}

\label{power logarithmic linked}

Another example belonging to the logarithmic linked exposure family is obtained by choosing the power function
\begin{equation*}
  b(z)=\left(1-\frac{z}{\alpha}\right)^\delta + a,
\end{equation*}
for $\alpha>1$, $\delta>1$ and a sufficiently large parameter $a > 0$. This is a smooth and strictly convex curve on $[0,1]$ with $b(1)=(1-1/\alpha)^\delta +a < 1+a = b(0)$. The first and second derivatives for $z\in [0,1]$ are
given by
\begin{equation*}
  b'(z)=-\frac{\delta}{\alpha}\left(1-\frac{z}{\alpha}\right)^{\delta-1}~<~0
  \qquad \text{ and } \qquad
   b''(z)=\frac{\delta(\delta-1)}{\alpha^2}\left(1-\frac{z}{\alpha}\right)^{\delta-2}~>~0.
\end{equation*}
In order to achieve $b''(z)b(z) - b'(z)^2 \geq 0$ for all $z \in [0,1]$, which is a necessary condition due to Proposition~\ref{prop exp curve 1}, the parameter $a$ has to satisfy
\begin{equation}\label{condition on a}
  a > \max_{z \in [0,1]} \left(\frac{b'(z)^2}{b''(z)}- \left(1-\frac{z}{\alpha}\right)^\delta\right).
\end{equation}
This provides in this example
\begin{equation}\label{side constraint a}
  a > \max_{z \in [0,1]}\left( \frac{1}{\delta-1}\left(1-\frac{z}{\alpha}\right)^{\delta}\right)~=~\frac{1}{\delta-1}.
  \end{equation}
With Corollary \ref{cor exp curve 1}, we then obtain as density for $z\in[0,1)$
\begin{equation}\label{general density}
  f_Z(z)=
\frac{a+1}{\alpha} \frac{\left(1-{z}/{\alpha}\right)^{\delta-2}}{\left(a+(1-{z}/{\alpha})^\delta\right)^2}
 \left(a(\delta-1)- \left(1-{z}/{\alpha}\right)^{\delta}\right)       
            ,
\end{equation}
with point mass in $z=1$ equal to
\begin{equation*}
  p = \left(1-\frac{1}{\alpha}\right)^{\delta-1}\frac{a+1}{a+(1-{1}/{\alpha})^\delta},
\end{equation*}
and mean
\begin{equation*}
  \E[Z] = \frac{\alpha(a+1)}{\delta}\, \log\left(\frac{a+1}{a+(1-{1}/{\alpha})^\delta}\right).
\end{equation*}
The derivative of the density $f_Z$ is given by
\begin{equation}\label{this is the derivative}
f'_Z(z)=
  -\frac{a+1}{\alpha^2}\, \frac{\left[a^2(\delta-1)(\delta-2)-a(\delta-1)(\delta+4)(1-{z}/{\alpha})^{\delta}
 +2(1-{z}/{\alpha})^{2\delta} \right]}
      {(a+(1-{z}/{\alpha})^\delta)^3\,(1-{z}/{\alpha})^{-\delta+3}}.
\end{equation}

\begin{lemma}\label{unimodal is in question}
  The power logarithmic linked exposure example with the above parameters leads to a well-defined distribution. The density $f_Z$ of this power logarithmic linked exposure example given in \eqref{general density} can only be unimodal on $[0,1)$ if $\delta >2$.
\end{lemma}

\begin{figure}[htb!]
    \centering
    \includegraphics[scale=0.7]{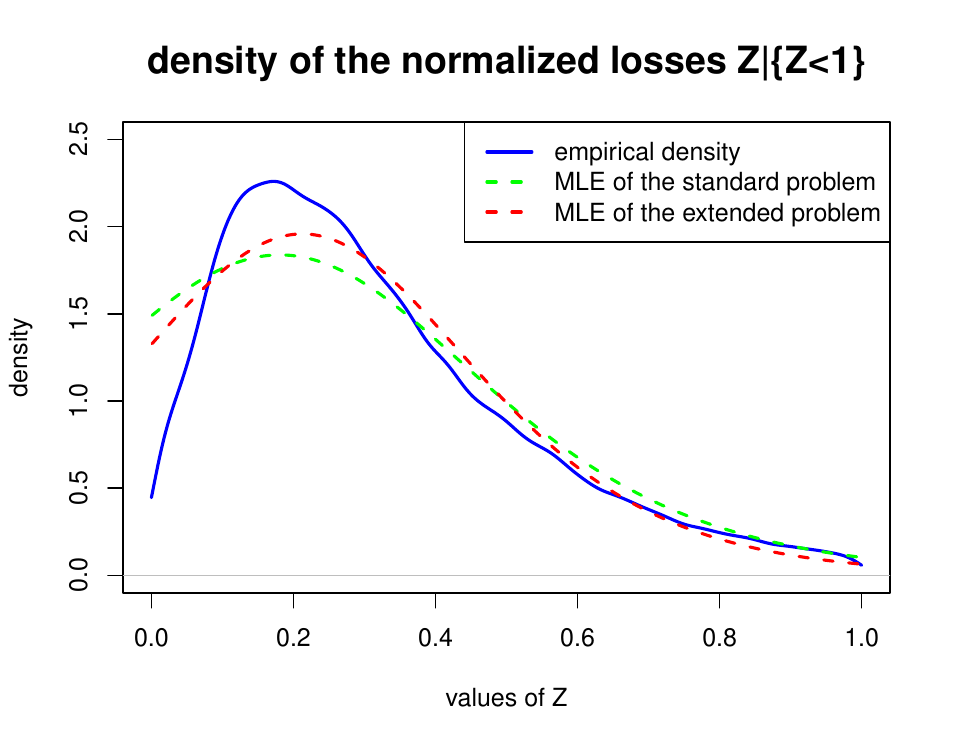}
    \caption{Power logarithmic linked exposure example: densities of the random variable $Z|_{\{Z<1\}}$.}
    \label{Fig:Power Log}
\end{figure}

Therefore, we restrict to the domain
\begin{equation*}
    \Theta = \Big\{ \alpha > 1, \delta > 2, a> \frac{1}{\delta-1} \Big\},
\end{equation*}
in order to find unimodal solutions to the MLE of the standard and extended maximization problems described in \eqref{log lik 1} and \eqref{log lik 2}. The results displayed in Figure \ref{Fig:Power Log} are very similar to the ones obtained for the MBBEFD example. This can be confirmed by comparing Table \ref{Tab:MBBEFD} and Table \ref{Tab:Power Log}, where most of the values coincide, although they are actually different if we look at digits after the decimal point. We conclude that this power logarithmic linked example does not improve the fit provided by the MBBEFD example, although this model allows for skewness. 

\begin{table}[htb!]
    \centering
    \begin{tabular}{ l c c c c c}
      \toprule
       & Point {m}ass & Mean & $\ell_{Z|_{\{Z<1\}}}$ & $\ell_Z$ & AIC \\
      \midrule
      Empirical density (Blue) & 0.034 & 0.339 & - & - & - \\
      MLE of the standard problem (Green) & 0.020 & 0.337 &32 680 &13 475 & -26 945 \\
      MLE of the extended problem (Red) & 0.034 & 0.336 &33 257 &14 587 & -29 166 \\
      \bottomrule
    \end{tabular}
    \caption{Power logarithmic linked exposure example: results.}
    \label{Tab:Power Log} 
\end{table}

\subsection{Sine logarithmic linked exposure example}

\label{sine logarithmic linked}

A third example belonging to the logarithmic linked exposure family is given by the sine function
\begin{equation*}
  b(z)= \sin(\alpha z + \beta) + a,
\end{equation*}
for $\beta \in \left(-\frac{\pi}{2},0 \right), \alpha \in \left(0, \frac{\pi}{2} - \beta \right)$ and $ - \sin(\beta) < a < - 1/\sin(\beta)$. This is a smooth curve on $[0,1]$ with $b(0)=\sin(\beta) + a < \sin(\alpha + \beta) + a = b(1)$. The first and second derivatives for $z\in [0,1]$ are
given by
\begin{equation*}
  b'(z)= \alpha \cos(\alpha z + \beta) ~>~0
  \qquad \text{ and } \qquad
   b''(z)=- \alpha^2 \sin(\alpha z + \beta).
\end{equation*}
Note that in this case, the function $b$ is in general neither concave, nor convex on the entire interval $[0,1]$. We claim that $b''(z)b(z) - b'(z)^2 \geq 0$ holds for all $z \in [0,1]$, which is a necessary condition in order to obtain a distribution function due to Proposition \ref{prop exp curve 1}. Moreover, the density for $z\in[0,1)$ reads as
\begin{equation}
    \label{density sine}
  f_Z(z)= \frac{\sin(\beta)+a}{\cos(\beta)} \, \frac{ \alpha \left[1+a \sin(\alpha z + \beta)\right]}{\left[\sin(\alpha z + \beta)+a\right]^2},  
\end{equation}
with point mass in $z=1$
\begin{equation*}
  p = \frac{\cos(\alpha + \beta)}{\cos(\beta)} \, \frac{\sin(\beta)+a}{\sin(\alpha +\beta)+a},
\end{equation*}
and mean
\begin{equation*}
  \E[Z] = - \frac{\sin(\beta) + a}{ \alpha \cos(\beta)}\, \log\left(\frac{\sin(\beta) + a}{\sin(\alpha + \beta) + a}\right).
\end{equation*}
The derivative of the density $f_Z$ is given by
\begin{equation} \label{sinusoidal derivative}
f'_Z(z)=
  -\frac{\sin(\beta) + a}{\cos(\beta)}\, \frac{\alpha^{2} \cos(\alpha z + \beta) \left[-a^2 +a \sin(\alpha z + \beta) + 2 \right]}
      {\left[\sin(\alpha z + \beta)+a\right]^3}.
\end{equation}

\begin{lemma}\label{sine distribution and unimodal}
  The sine logarithmic linked exposure example with the above parameters leads to a well-defined distribution. Moreover, the density $f_Z$ given in \eqref{density sine} can only be unimodal on $[0,1)$ if $1 \leq a \leq 2$.
\end{lemma}

\begin{figure}[htb!]
    \centering
    \includegraphics[scale=0.7]{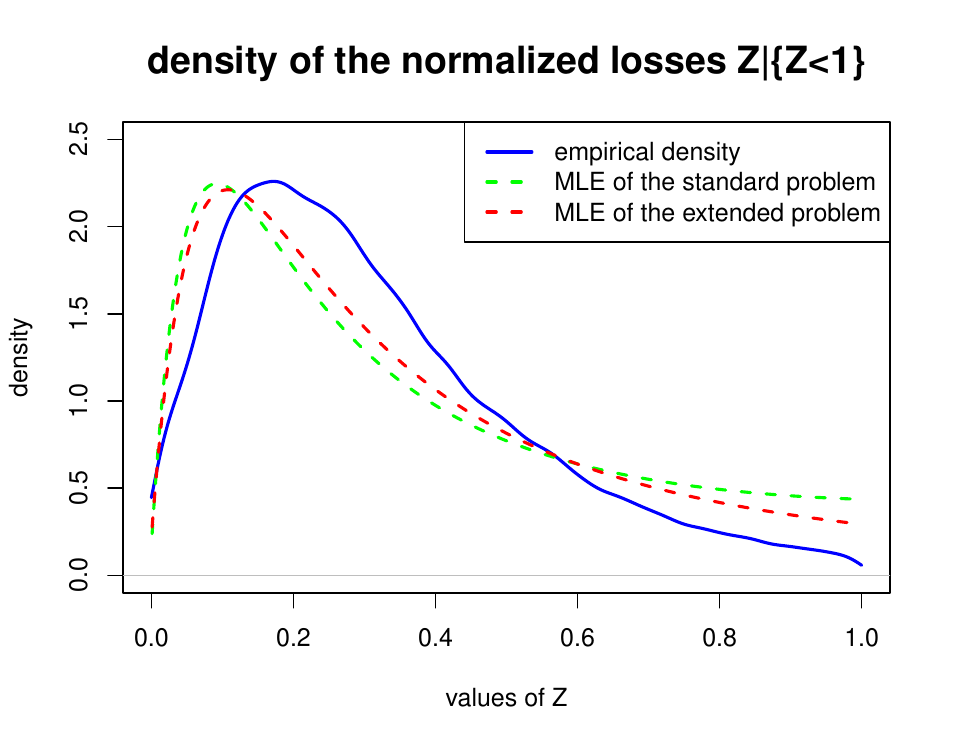}
    \caption{Sine logarithmic linked exposure example: densities of the random variable $Z|_{\{Z<1\}}$.}
    \label{Fig:Sine Log}
\end{figure}

\begin{table}[htb!]
    \centering
    \begin{tabular}{ l c c c c c}
      \toprule
       & Point mass & Mean & $\ell_{Z|_{\{Z<1\}}}$ & $\ell_Z$ & AIC \\
      \midrule
      Empirical density (Blue) & 0.034 & 0.339 & - & - & - \\
      MLE of the standard problem (Green) & 0.040 & 0.380 &26 907 &8 164 & -16 323 \\
      MLE of the extended problem (Red) & 0.034 & 0.362 &30 453 &11 783 & -23 558 \\
      \bottomrule
    \end{tabular}
    \caption{Sine logarithmic linked exposure example: results.}
    \label{Tab:Sine Log} 
\end{table}

Thus, we use the domain
\begin{equation*}
    \Theta = \left\{ \beta \in \left(-\frac{\pi}{2},0 \right), \alpha \in \left(0, \frac{\pi}{2} - \beta \right), 1 < a < \min \left(- \frac{1}{\sin(\beta)}, 2  \right) \right\},
\end{equation*}
in order to find unimodal solutions to the standard and extended maximization problems described in \eqref{log lik 1} and \eqref{log lik 2}. Similarly as for the previous examples, we obtain the results displayed in Figure \ref{Fig:Sine Log} and Table \ref{Tab:Sine Log}. Although this example manages to produce rather skewed densities, see Figure \ref{Fig:Sine Log}, the fit does not seem to be accurate for this data, i.e., it is less accurate than the first examples, and we clearly prefer the previous models.

\subsection{Quadratic exponentially linked exposure example}
\label{Quadratic exponential linked exposure example}
Let us now treat an example belonging to the exponentially linked exposure family, and given by the quadratic inner function
\begin{equation*}
  b(z)= \alpha z^2 + \beta z,
\end{equation*}
for $\alpha < 0$ and $\beta < - \sqrt{-2 \alpha}$. This is a smooth curve on $[0,1]$ with $b(0)=0 > \alpha + \beta = b(1)$. The first and second derivatives for $z\in [0,1]$ are
given by
\begin{equation*}
  b'(z)= 2 \alpha z + \beta ~<~0
  \qquad \text{ and } \qquad
   b''(z)= 2 \alpha ~<~0.
\end{equation*}
This function $b$ is smooth and strictly concave on $[0,1]$. We claim that $b''(z)+b'(z)^2\geq0$ holds for all $z \in [0,1]$, which is a necessary condition in order to obtain a distribution function due to Proposition \ref{prop exp curve 2}. Moreover, the density for $z\in[0,1)$ is given by
\begin{equation}
    \label{density quadratic}
  f_Z(z) = - \frac{e^{\alpha z^2 + \beta z} \left[4 \alpha^2 z^2 + 4 \alpha \beta z + \beta^2 + 2 \alpha \right]}{\beta},
\end{equation}
with point mass in $z=1$
\begin{equation*}
  p = \frac{e^{\alpha + \beta } \, [2 \alpha  + \beta]}{\beta},
\end{equation*}
and mean
\begin{equation*}
  \E[Z] = \frac{e^{\alpha + \beta } - 1}{\beta}.
\end{equation*}
The derivative of the density $f_Z$ is given by
\begin{equation} \label{quadratic derivative}
f'_Z(z)=
  - \frac{e^{\alpha z^2 + \beta z} \, (2 \alpha z + \beta)  \left[(2 \alpha z + \beta)^2 + 6 \alpha \right]}{\beta}.
\end{equation}

\begin{lemma}\label{quadratic distribution and unimodal}
  The quadratic exponentially linked exposure example with the above parameters leads to a well-defined distribution. Moreover, the density $f_Z$ given in \eqref{density quadratic} is unimodal on $[0,1)$ if and only if $- \sqrt{-6 \alpha} < \beta < -2 \alpha - \sqrt{-6 \alpha}$.
\end{lemma}

\begin{figure}[htb!]
    \centering
    \includegraphics[scale=0.7]{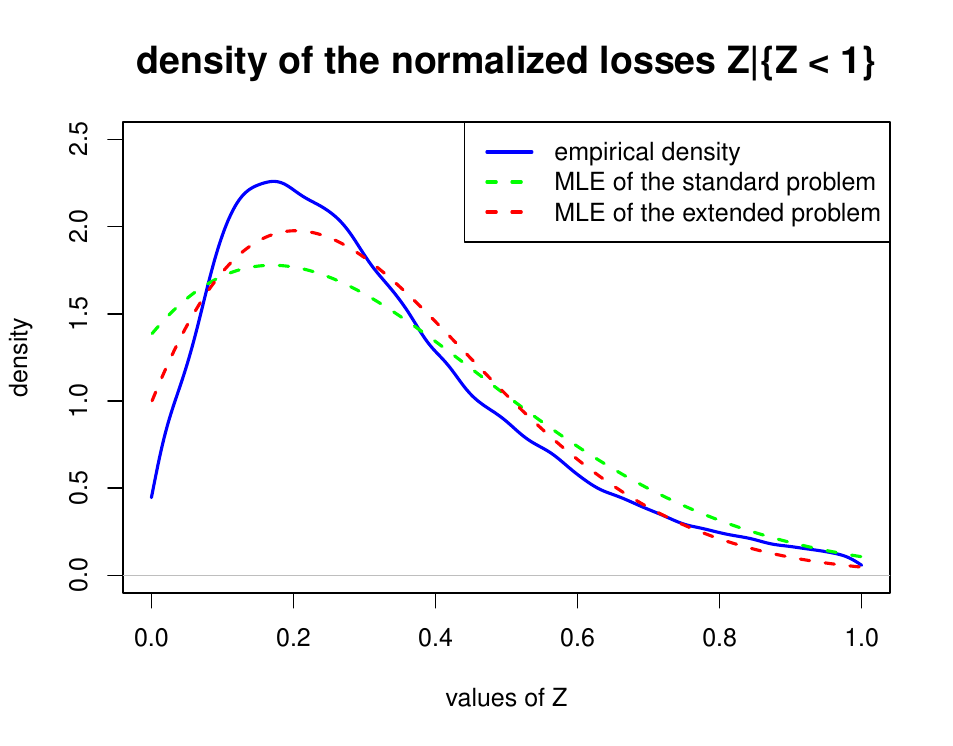}
    \caption{Quadratic exponentially linked exposure example: densities of the random variable $Z|_{\{Z<1\}}$.}
    \label{Fig:Quadratic Log}
\end{figure}

Thus, we restrict to the domain
\begin{equation*}
    \Theta = \Big\{ \alpha < 0, \beta \in \left(- \sqrt{-6 \alpha} \,, \min \left(- 2 \alpha - \sqrt{-6 \alpha}, - \sqrt{-2 \alpha} \right) \right) \Big\} ,
\end{equation*}
in order to find unimodal solutions to the standard and extended maximization problems described in \eqref{log lik 1} and \eqref{log lik 2}. Similarly as for the previous examples, we obtain the results shown in Figure \ref{Fig:Quadratic Log} and Table \ref{Tab:Quadratic Log}. Although a new family of exposure curves is used here, the results are close to the ones obtained with the power logarithmic linked exposure example or the MBBEFD example. In fact, in the extended problem, we obtain a slightly better model than in the MBBEFD class according to AIC. However, looking at Figure \ref{Fig:Quadratic Log}, this new model is again not satisfactory for this data.

\begin{table}[htb!]
    \centering
    \begin{tabular}{ l c c c c c}
      \toprule
       & Point mass & Mean & $\ell_{Z|_{\{Z<1\}}}$ & $\ell_Z$ & AIC \\
      \midrule
      Empirical density (Blue) & 0.034 & 0.339 & - & - & - \\
      MLE of the standard problem (Green) & 0.016 & 0.345 &32 101 &12 425 &-24 846 \\
      MLE of the extended problem    (Red) & 0.034 & 0.341 &33 397 &14 726 &-29 447 \\
      \bottomrule
    \end{tabular}
    \caption{Quadratic exponentially linked exposure example: results.}
    \label{Tab:Quadratic Log} 
\end{table}

\subsection{Power exponentially linked exposure example}

\label{power exponential link}

\begin{figure}[htb!]
    \centering
    \includegraphics[scale=0.7]{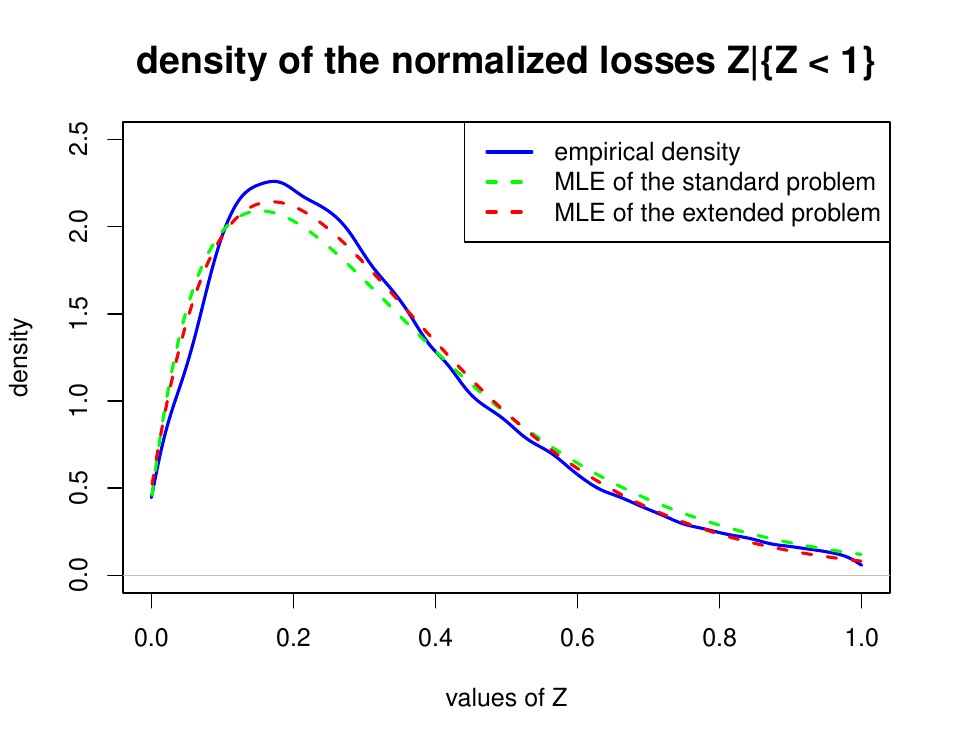}
    \caption{Power exponentially linked exposure example: densities of the random variable $Z|_{\{Z<1\}}$.}
    \label{Fig:Power exp}
\end{figure}

We consider a final example belonging to the exponentially linked exposure family. This example is a bit more difficult in handling, but it provides the best results for our dataset. Choose the power function
\begin{equation*}
  b(z)= \epsilon (z+\delta)^\alpha - \beta z,
\end{equation*}
for $\alpha \in (1,2), \delta > 0, \epsilon < 0$ and $\beta > \epsilon \alpha \delta^{\alpha-1} + \sqrt{-\epsilon \alpha (\alpha-1) \delta^{\alpha-2}}$. This is a smooth curve on $[0,1]$ with $b(0)=\epsilon \delta^\alpha > \epsilon (1+\delta)^\alpha - \beta = b(1)$. The first and second derivatives for $z\in [0,1]$ are
given by
\begin{equation*}
  b'(z)= \alpha \epsilon (z+\delta)^{\alpha-1} - \beta ~<~0
  \qquad \text{ and } \qquad
   b''(z)= \alpha (\alpha-1) \epsilon (z+\delta)^{\alpha-2} ~<~0.
\end{equation*}
This function $b$ is smooth and strictly concave on $[0,1]$. We claim that $b''(z) + b'(z)^2 \geq 0$ holds for all $z \in [0,1]$, which is a necessary condition in order to obtain a distribution function due to Proposition \ref{prop exp curve 2}. Moreover, the density for $z\in[0,1)$ reads as
\begin{equation*}
  f_Z(z) = - \frac{e^{\epsilon (z+\delta)^\alpha - \beta z} \left[(\alpha \epsilon (z+\delta)^{\alpha-1} - \beta)^2 + \alpha (\alpha-1) \epsilon (z+\delta)^{\alpha-2}\right]}{e^{\epsilon\delta^\alpha}  \left[\alpha \epsilon \delta ^{\alpha-1} - \beta \right]},
\end{equation*}
with point mass in $z=1$
\begin{equation*}
  p = \frac{e^{\epsilon (1+\delta)^\alpha - \beta} \, [\alpha \epsilon (1+\delta)^{\alpha-1} - \beta]}{e^{\epsilon \delta^\alpha} \, [\alpha \epsilon \delta^{\alpha-1} - \beta]},
\end{equation*}
and mean
\begin{equation*}
  \E[Z] = \frac{e^{\epsilon \left[(1+\delta)^\alpha - \delta^\alpha \right] - \beta}- 1}{\alpha \epsilon \delta^{\alpha-1} - \beta}.
\end{equation*}
The derivative of the density $f_Z$ can be obtained using \eqref{derivative exp curve 2}. It does however not allow to explicitly characterize the extrema of the density in this example. Nevertheless, we observe in Figure \ref{Fig:Power exp} that this example contains unimodal densities.

\begin{lemma}\label{power exp distribution and unimodal}
The power exponentially linked exposure example with the above parameters leads to a well-defined distribution.
\end{lemma}

Thus, we restrict to the domain
\begin{equation*}
    \Theta = \Big\{ \alpha \in (1,2), \delta > 0, \epsilon < 0, \beta > \epsilon \alpha \delta^{\alpha-1} + \sqrt{-\epsilon \alpha (\alpha-1) \delta^{\alpha-2}} \Big\},
\end{equation*}

in order to find solutions to the standard and extended maximization problems described in \eqref{log lik 1} and \eqref{log lik 2}. Similarly as for the previous examples, we obtain the results shown in Figure \ref{Fig:Power exp} and Table \ref{Tab:Power exp}. This time, the fit seems much better. This is especially the case for the red curve, which represents the solution of the extended maximization problem. The tail behavior on the right and left ends seems adequate in contrast to the plots presented in the previous examples. This observation is confirmed by the AIC scores attained here, which are lower than the AIC scores of all other examples, i.e., we give clear preference to this last example for this data.

\begin{table}[htb!]
    \centering
    \begin{tabular}{ l c c c c c}
      \toprule
       & Point mass & Mean & $\ell_{Z|_{\{Z<1\}}}$ & $\ell_Z$ & AIC \\
      \midrule
      Empirical density (Blue) & 0.034 & 0.339 & - & - & - \\
      Standard MLE density (Green) & 0.025 & 0.341 &34 068 &15 199 &-30 390 \\
      Flexible MLE density (Red) & 0.034 & 0.339 &34 402 &15 731 &-31 453 \\
      \bottomrule
    \end{tabular}
    \caption{Power exponentially linked exposure example: results.}
    \label{Tab:Power exp} 
\end{table}

\subsection{Comparison with the log-normal and gamma distribution}

\label{comparison}

\begin{figure}[htb!]
    \centering
    \includegraphics[scale=0.71]{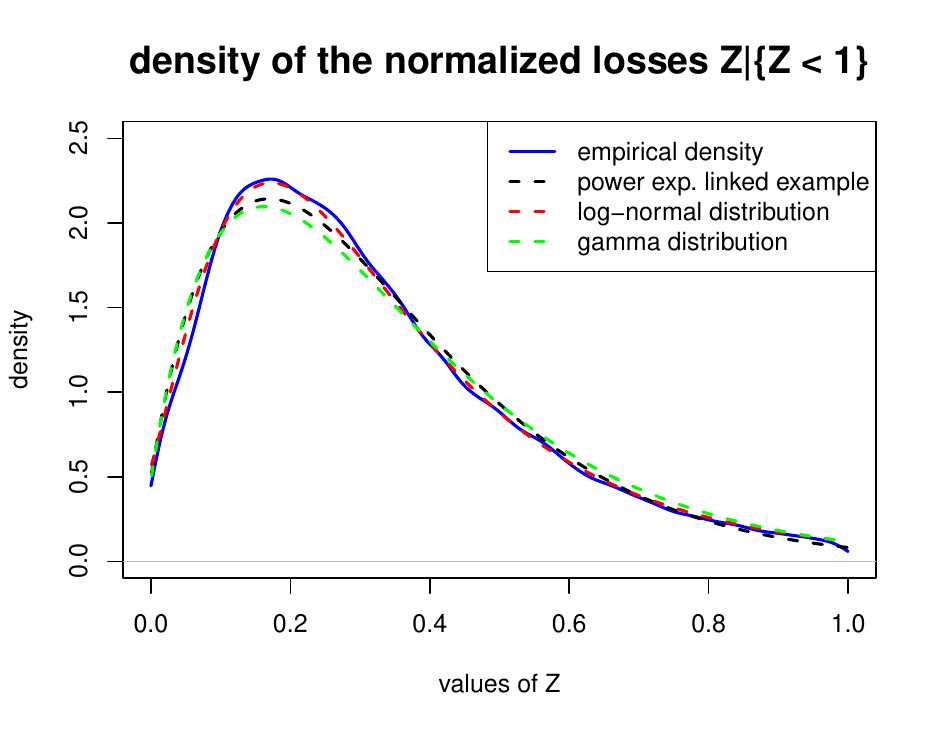}
    \caption{Comparison with the gamma and the log-normal distribution: densities of the random variable $Z|_{\{Z<1\}}$.}
    \label{Fig:Comparison}
\end{figure}

For this real dataset example, we performed an MLE estimation without taking into account any covariates. In that case, performing MLE using lower-truncated and right-censored log-normal and gamma models is feasible. Our aim is to compare the extended fit of the power exponentially linked exposure example to the fit of these classical distributions.
For this, we first consider the normalized claim $Z$ and write
\begin{equation*}
    Z = \frac{1}{M} \min \left\{ \left(X-d \right)_{+}, M \right\} \, | \, X > d = \left. \min \left\{ \left(\frac{X}{M}-\frac{d}{M}\right)_{+}, 1 \right\} \, \right| \, \frac{X}{M} > \frac{d}{M},
\end{equation*}
where $X$ is assumed to have a log-normal and gamma distribution, respectively. For an arbitrary $M > 0$, both distributions share the following scaling property
\begin{equation*}
    X \sim \textrm{LN}(\mu, \sigma^2) \implies \frac{X}{M} \sim \textrm{LN}(\mu - \log(M), \sigma^2) \quad \textrm{and} \quad X \sim \Gamma(\gamma, c) \implies \frac{X}{M} \sim \Gamma(\gamma, cM),
\end{equation*}
where $\mu \in \R$ and $\sigma^2 > 0$ are the parameters of a log-normal distribution, and $\gamma > 0$ and $c > 0$ are the shape and scale parameters of a gamma distribution. Let us denote by $\theta$ the parameters of the log-normal, respectively, gamma distribution. Due to the above scaling property and by setting $\tilde{d} = d/M$, we can maximize without loss of generality the log-likelihood function
\begin{equation}
    \label{MLE gamma lognormal}
    (\theta, \tilde{d}) \mapsto \ell_{Z}(\theta, \tilde{d}) = \log(f^{(\theta, \tilde{d})}(Z)) \mathds{1}_{\{Z<1\}} + \log(p^{(\theta, \tilde{d})}) \mathds{1}_{\{Z=1\}},
\end{equation}
where the absolutely continuous density $f^{(\theta, \tilde{d})}$ is given by
\begin{equation*}
    f^{(\theta, \tilde{d})}(y) = \frac{f_X(\tilde{d}+y)}{1-F_X(\tilde{d})},
\end{equation*}
for $y \in [0,1)$, $\tilde{d} \in (0,\infty)$, and where the point mass $p^{(\theta, \tilde{d})}$ is given by 
\begin{equation*}
    p^{(\theta, \tilde{d})} = \frac{1-F_X(\tilde{d}+1)}{1-F_X(\tilde{d})},
\end{equation*}
see \eqref{lower-truncated and right-censored gamma density} and \eqref{log-likelihood right-censoring}. For both distributions, the number of parameters is thus equal to three and the results of these MLE fits are given in Figure \ref{Fig:Comparison} and Table \ref{Tab:Comparison}. Compared to the extended MLE of the power exponentially linked example, the fit on the absolutely continuous density seems slightly worse for the gamma model, whereas the fit using the log-normal model seems to be accurate for this dataset when looking at Figure \ref{Fig:Comparison}. These observations are confirmed in Table \ref{Tab:Comparison}. On the one hand, the gamma model fails to obtain a suitable value for the point mass, and on the other hand, the values obtained for the point mass and the mean are close to the empirical values for the log-normal model, which also achieves the lowest AIC score. Therefore we give preference to the log-normal model over the other models for this dataset.

\begin{table}[htb!]
    \centering
    \begin{tabular}{ l c c c c c}
      \toprule
       & Point mass & Mean & $\ell_{Z|_{\{Z<1\}}}$ & $\ell_Z$ & AIC \\
      \midrule
      Empirical density (Blue) & 0.034 & 0.339 & - & - & - \\
      \specialcell{Flexible MLE density of the power \\ exponentially linked example (Black)} & 0.034 & 0.339 &34 402 &15 731 &-31 453 \\
      \specialcell{MLE density of the log-normal \\ distribution (Red)} & 0.031 & 0.340 &34 640 &15 956 &-31 907 \\
      \specialcell{MLE density of the gamma \\ distribution (Green)} & 0.024 & 0.341 &34 236 &15 355 &-30 704 \\
      \bottomrule
    \end{tabular}
    \caption{Comparison with the gamma and the log-normal distribution: results.}
    \label{Tab:Comparison}
\end{table}

Thus, it seems that the log-normal distribution has a better performance than our examples from the Bernegger class. However, we would like to point out that for fitting the gamma and the log-normal models, we do not work with the original deductible $d$ and maximal cover $M$ of the product since we optimize over $\tilde{d} = d/M$, see \eqref{MLE gamma lognormal}. This means that we do not assume the log-normal and the gamma distribution to fit the total financial losses, but we allow for arbitrary scaling. We also remark that fitting the gamma and log-normal models under the original deductible $d$ and maximal cover $M$ does not lead to competitive models for this dataset. Of course, it is then unclear how one could derive the distribution of the total financial loss from the distribution of the observed insurance claims. We show in the next section that such a derivation is in general not unique and implies to make possibly wrong assumptions on the distribution of the total financial loss. 
\section{Changes in the deductible or the maximal cover}

\label{Extrapolation}

The primary objective of modeling lower-truncated and right-censored insurance claims is to fit the size of observed claims, or equivalently, the size of financial losses falling in the interval $(d, d+M)$. However, the insurer might also be interested in understanding how a change in the deductible and/or in the maximal cover influences the expected claim size. Let us denote by $\tilde{d}$ the new deductible and by $\tilde{M}$ the new maximal cover. Two cases may arise.

If the deductible increases or the maximal cover decreases such that $(\tilde{d}, \tilde{d}+\tilde{M}) \subseteq (d, d+M)$, the insurer can use his current claims observations in order to derive a new model for the new range of interest. However, in the case where $(\tilde{d}, \tilde{d}+\tilde{M}) \not\subseteq (d, d+M)$, e.g., when the deductible decreases and the maximal cover increases, an extrapolation based on the distribution of the observed total financial losses being in $(d, d+M)$ has to be made in order to obtain a candidate for the distribution on the new range of interest. We discuss in this section how the insurer can evaluate the new expected claim size after a change in the deductible or the maximal cover by treating separately the two above cases.

\subsection{Increase in the deductible or decrease in the maximal cover}
As in the previous sections, we denote the total financial loss by $X$, whereas the insurance claim is denoted by $Y$, see \eqref{claim transform}. Furthermore, we define the normalized insurance claim by
\begin{equation*}
    Z = \frac{1}{M} \min \{(X-d)_{+}, M \} \, | \, Z > d.  
\end{equation*}
After an increase in the deductible or a decrease in the maximal cover such that the new range of interest satisfies $(\tilde{d}, \tilde{d}+\tilde{M}) \subseteq (d, d+M)$, the new normalized insurance claim becomes
\begin{equation*}
    \begin{split}
    \tilde{Z} &= \frac{1}{\tilde{M}} \min \left\{(X-\tilde{d})_{+}, \tilde{M} \right\} \, | \, X > \tilde{d} \\
    &= \frac{M}{\tilde{M}} \left. \min \left\{\left(Z-\frac{\tilde{d}-d}{M}\right)_{+}, \frac{\tilde{M}}{M} \right\} \, \right|  \, Z > \frac{\tilde{d}-d}{M}.
    \end{split}
\end{equation*}
That is, the distribution of $\tilde{Z}$ is equal to a scaled lower-truncated and right-censored distribution of $Z$, where the lower-truncation point is given by $\bar{d} = (\tilde{d}-d)/M$ and the right-censoring point is $\bar{M} = \tilde{M}/M$ (after subtracting the lower-truncation). The next result shows that if $Z$ belongs to the Bernegger class, then $\tilde{Z}$ also belongs to the Bernegger class. In other words, the Bernegger class is closed under lower-truncation and right-censoring.
\begin{prop}
    \label{Bernegger closed}
    Let $Z$ be a member of the Bernegger class with exposure curve $G$ and let $0 \leq \bar{d} , \, \bar{M} \leq 1$ such that $\bar{d} + \bar{M} \leq 1$. Moreover, define the scaled lower-truncated and right-censored random variable
    \begin{equation*}
        \tilde{Z} = \frac{1}{\bar{M}}\min \{(Z-\bar{d})_{+}, \bar{M} \} \, | \, Z > \bar{d}.    
    \end{equation*}
    This random variable belongs again to the Bernegger class and its exposure curve is given by
    \begin{equation*}
        \tilde{G} : [0,1] \to [0,1], \; z \mapsto \frac{G(\bar{d}+z\bar{M})}{G(\bar{d}+\bar{M})}.
    \end{equation*}
\end{prop}
We point out that in the case where an increase in the deductible and/or a decrease in the maximal cover leads to a new interval satisfying $(\tilde{d}, \tilde{d}+\tilde{M}) \not\subseteq (d, d+M)$, this last result does not apply since either the lower-truncation point $\bar{d}= (\tilde{d}-d)/M$ is negative or the sum $\bar{d} + \bar{M} =(\tilde{d}-d+\tilde{M})/M$ exceeds $1$. This situation has then to be handled by performing some extrapolation on the observed part of the total financial loss distribution as in the case where the deductible decreases or the maximal cover increases, see next section.

\subsection{Decrease in the deductible or increase in the maximal cover}

The goal of this last section is to treat the case where the new deductible and the new maximal cover satisfy $(\tilde{d}, \tilde{d}+\tilde{M}) \not\subseteq (d, d+M)$. This typically happens when the deductible $d$ decreases or when the maximal cover $M$ increases, but also in some cases when $d$ increases and $M$ decreases, as discussed above. Since the insurer only has full knowledge on the distribution of the total final loss on the interval $(d, d+M)$, he will have to perform an extrapolation of the observed density in order to obtain a candidate for the distribution on the new range of interest.

\begin{figure}[htb!]
\begin{center}
\begin{minipage}[t]{0.45\textwidth}
\begin{center}
\includegraphics[width=\textwidth]{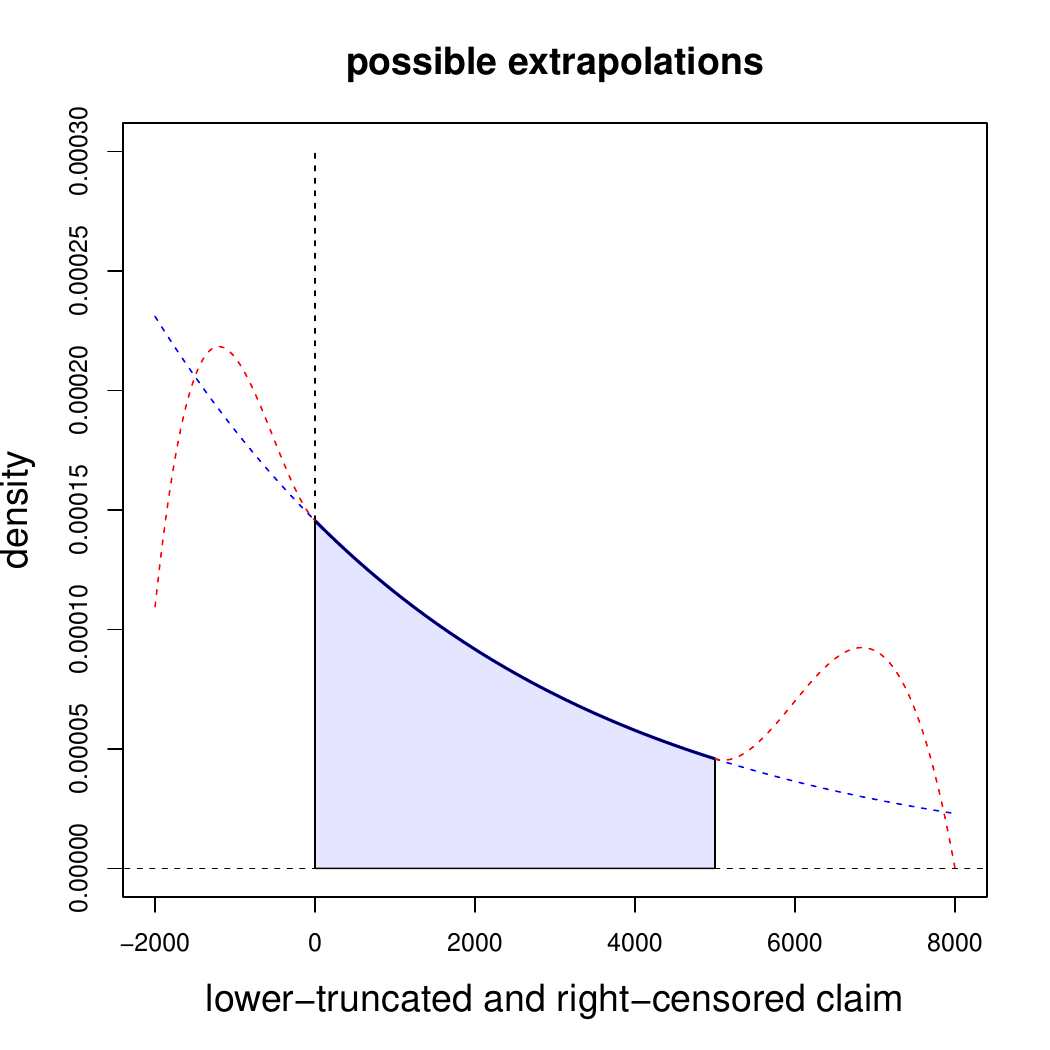}
\end{center}
\end{minipage}
\end{center}
\vspace{-.7cm}
\caption{Two possible extrapolations of the density of a lower-truncated and right-censored exponential random variable with $d=2000$ and $M=5000$.}
\label{Fig:Extrapolation}
\end{figure}

In Example \ref{Exponential distribution}, we showed that a scaled lower-truncated and right-censored exponential random variable belongs to the Bernegger class. Using this example, we show in Figure \ref{Fig:Extrapolation} that extrapolating a density on a given interval to a larger interval leads to infinitely many possible candidates. For this, we first plot in blue the density of an exponential random variable $X \sim \textrm{Exp}(\lambda)$ for $\lambda = \log(2)/3000$. Then, we set the deductible to $d = 2000$ and the maximal cover $M = 5000$. The blue area in Figure \ref{Fig:Extrapolation} corresponds to the observable part of the density of the total financial loss of the insurer. Let us denote by $p_{-}$ the probability of $X$ being smaller than $2000$ and by $p_{+}$ the probability of $X$ being larger than $7000$. We plot in red another possible extrapolation of the observed density such that the density of the total financial loss under this new extrapolation is continuously differentiable and such that the area under the red curve at the left and right ends is equal to $p_{-}$ and $p_{+}$, respectively. Using the observations falling in the interval $(d, d+M)$ as well as the point mass in $d+M$ indicating the proportion exceeding this threshold, the insurer may choose either one of the two above extrapolations and, actually, the number of possible candidates is in general infinite. Therefore, we point out that in the case where $(\tilde{d}, \tilde{d}+\tilde{M}) \not\subseteq (d, d+M)$, the insurer cannot derive the distribution of the total financial loss on this new range without making additional assumptions on the random variable~$X$. Using the two-parameter logistic distribution of Example \ref{Logistic distribution}, the next example shows a possible extrapolation.

\begin{example}[Lower-truncated and right-censored two-parameter logistic distribution]

\normalfont \label{last Logistic distribution}
    Let us assume that $X$ follows a two-parameter logistic distribution as in Example \ref{Logistic distribution}. Moreover, let $d \in \R$ and $M > 0$. By Corollary \ref{cor exp curve 1}, the normalized lower-truncated and right-censored random variable $Z$ in \eqref{lower-truncated and right-censored logistic} has as density
\begin{equation}
    \label{f_Z logistic}
        f_Z(z) = \frac{M}{\sigma} \left( 1+ e^{(d-\mu)/\sigma} \right)\frac{e^{(zM+d-\mu)/\sigma}}{\left(1+e^{(zM+d-\mu)/\sigma} \right)^2},
\end{equation}
for $z \in [0,1)$, with a point mass in $1$ equal to
\begin{equation*}
    p = \frac{1+ e^{(d-\mu)/\sigma}}{1+ e^{(M+d-\mu)/\sigma}}.
\end{equation*}
Suppose now that we know the values of the $d$, $M$ and $p$ and that we want to retrieve the unknown distribution of the random variable $X$. This distribution is related to the distribution of $Z$ by
\begin{equation}
 \label{dist of X Z logistic}
    F_{Z}(z) = \frac{F_X(d+zM) - F_X(d)}{1-F_X(d)}
\,\mathds{1}_{\{z \in [0,1) \}} + \mathds{1}_{\{z =1\}}.
\end{equation}
At this point, we have to make the following (possibly wrong) assumptions. We first assume that the random variable $X$ is absolutely continuous and that the support of its density is the whole real line. Moreover, we assume that
\begin{equation}
    \label{cst assumption}
    f_X(z) = C f_Z \left(\frac{z-d}{M} \right),
\end{equation}
for $z \in \R$ and where $C$ is a normalizing constant such that $f_X$ becomes a density on $\R$. Note that in this last assumption, $f_Z$ is now seen as a function defined on the whole real line. In general, it might not be possible to obtain a well-defined function by extending the domain of $f_Z$, which is a priori only defined for $z \in [0,1)$, see \eqref{f_Z logistic}. In this case, however, extending the domain of the density $f_Z$ to the whole real line leads to a well-defined function. Note that assumption \eqref{cst assumption} might seem natural in view of \eqref{dist of X Z logistic}. Our above assumptions give us again the original density in this case, namely,
\begin{equation*}
    f_{X}(z) = \frac{e^{(z-\mu)/\sigma}}{\sigma \left( 1 + e^{(z-\mu)/\sigma} \right)^2}, \quad \textrm{for} \, -\infty < z < \infty, \, -\infty < \mu < \infty, \, \sigma > 0.
\end{equation*}
\EndExample
\end{example}

This example shows how we can start from a member of the Bernegger class in order to obtain a candidate for the original distribution. Under some assumptions, we managed to retrieve the original distribution in this case. If we assumed however that the support of the density $f_X$ is the positive real line or that $X$ is not an absolutely continuous random variable but rather has some point masses at selected locations, we would have ended with a wrong candidate for the original distribution. Of course, the point mass in $1$ can help us in order to verify whether the made assumptions are plausible but even with this information, the number of possible candidates for the original distribution is in general infinite. That is, we point out again that it is in general impossible to retrieve the original distribution by performing some extrapolation based on the lower-truncated and right-censored distribution.

\section{Conclusion}
\label{sec: Conclusion}
Most classical statistical models are not directly suited to model lower-truncated and right-censored claims in general insurance since they lead to problems that are not easily analytically tractable. Bernegger introduced in \cite{Bernegger} the MBBEFD class of distributions that can model such claims using a distribution function, an absolutely continuous density, and a point mass that are all of closed form. This class was introduced in the reinsurance literature, where densities are typically monotonically decreasing.

In general insurance, however, we are mainly interested in unimodal skewed densities. Therefore, we extended the MBBEFD class to a much bigger family of distributions that we called the Bernegger class. By starting from the properties of an exposure curve, we introduced two subfamilies, namely the logarithmic and exponentially linked exposure families. Through various examples, we used the full tractability and flexibility of the Bernegger class in order to fit parameters to general insurance claims using maximum likelihood estimations. It turned out that this large class of distributions contains models allowing to obtain a suitable approximation for the distribution of the used dataset, and in general, we have a rich family of unimodal and skewed densities within the Bernegger class. This class of distribution allows to model lower-truncated and right-censored random variables without making any assumption on the original unobserved distribution. We showed further that it is in general impossible to obtain a unique extrapolation from the observable lower-truncated and right-censored distribution in order to derive the full distribution of the original random variable. That is, from an actuarial perspective, it is not possible to know the distribution of the total financial loss from the insurance claims observations.

Going forward, it will be interesting to further characterize and classify the members of the Bernegger class based on different properties. Of course, this might involve exploring other link functions. Another next step is to lift these models to regression models allowing for integrating fixed effects described by covariates. In this last setup, we point out that relying on numerical approximation for the distribution function in order to perform MLE is not possible for a large number of covariates.

\bigskip

{\bf Acknowledgments.} We kindly thank the referees for their useful remarks that have helped us to improve this manuscript.

\newpage

\bigskip

{\small 
\renewcommand{\baselinestretch}{.51}
}

\newpage

\appendix


\section{Proofs}
We prove all statements in this appendix.

\bigskip

{\Beweis {\bf Proof of Theorem \ref{Exposure curves properties}.}
    The function $F_Z$, as defined in \eqref{from exposure to dist}, is continuously differentiable on the interval $(0,1)$ by our assumptions on $G$. This means that a derivative exists, and is equal to 
    \begin{equation*}
        f_Z(z) = F_Z'(z) = - \frac{G''(z)}{G'(0)} \geq 0,
    \end{equation*}
    for $z \in [0,1)$. Since $G$ was assumed to be twice continuously differentiable, non-decreasing and concave, we obtain that $0 \leq G'(z) \leq G'(0)$ and $f_Z(z) \geq 0$ for all $z \in [0,1]$. This implies 
    \begin{equation*}
        0 = F_Z(0) \leq F_Z(z) \leq F_Z(1) = 1, \, \text{ for all } z \in [0,1].
    \end{equation*}
    Since $G'$ is continuous by assumption, we conclude from \eqref{from exposure to dist} that $F_Z$ is right-continuous, and hence, a distribution function on $[0,1]$. The point mass in 1 is then given by
    \begin{equation*}
        p = 1 - F_Z(1_{-}) = \frac{G'(1)}{G'(0)} \in [0,1],
    \end{equation*}
    and the mean of $Z \sim F_Z$ is equal to
    \begin{equation*}
        \E[Z] = \int_0^1 1 - F_Z(s) \, ds = \int_0^1 \frac{G'(s)}{G'(0)} \, ds = \frac{G(1) - G(0)}{G'(0)} = \frac{1}{G'(0)}.
    \end{equation*}
    This proves the theorem.
\bigskip
\EndProof}

{\Beweis {\bf Proof of Lemma \ref{Mixture distributions}.}
First, it is clear that the function $G:[0,1] \rightarrow \R$ defined in \eqref{mixture exposure function} is an exposure curve. Moreover, we have by \eqref{from exposure to dist} that the distribution function $F_Z$ generated by $G$ satisfies

\begin{equation*}
    \begin{split}
        F_Z(z) &= \left(1- \frac{G'(z)}{G'(0)}\right) \mathds{1}_{\{z < 1\}} + \mathds{1}_{\{z = 1\}}  \\
        &= \left(1- \frac{\sum_{i=1}^n \alpha_i G_i'(z)}{\sum_{j=1}^n \alpha_j G_j'(0)}\right) \mathds{1}_{\{z < 1\}} + \mathds{1}_{\{z = 1\}} \\
        &= \left(1- \sum_{i=1}^n \frac{\alpha_i G_i'(0)}{\sum_{j=1}^n \alpha_j G_j'(0)} \frac{G_i'(z)}{G_i'(0)}\right) \mathds{1}_{\{z < 1\}} + \mathds{1}_{\{z = 1\}} \\
        &= \left(1- \sum_{i=1}^n w_i \frac{G_i'(z)}{G_i'(0)}\right) \mathds{1}_{\{z < 1\}} + \mathds{1}_{\{z = 1\}} \\
        &= \sum_{i=1}^n w_i \left(\left(1- \frac{G_i'(z)}{G_i'(0)}\right) \mathds{1}_{\{z < 1\}} + \mathds{1}_{\{z = 1\}} \right) = \sum_{i=1}^n w_i F_i(z),
    \end{split}
\end{equation*}
for $z \in [0,1]$, and where the elements
\begin{equation*}
    w_i = \frac{\alpha_i G_i'(0)}{\sum_{j=1}^n \alpha_j G_j'(0)}~\geq~0
\end{equation*}
are weights summing up to 1. The proof then follows similarly to Theorem \ref{Exposure curves properties}.
\bigskip
\EndProof}

{\Beweis {\bf Proof of Proposition \ref{analysis of modes}.}
We calculate the first derivative for $g>1$, $b>0$, $b\neq 1$ and $bg\neq 1$
\begin{eqnarray}\nonumber
  f'_{b,g}(z)&=& -
\frac{(g-1)(b-1)(\log(b))^2b^{1-z}}{\left((g-1)b^{1-z}+(1-bg)\right)^2}
+2\frac{(g-1)^2(b-1)(\log(b))^2b^{2-2z}}{\left((g-1)b^{1-z}+(1-bg)\right)^3}
\\&=&\nonumber
\frac{(g-1)(b-1)(\log(b))^2b^{1-z}}{\left((g-1)b^{1-z}+(1-bg)\right)^3}
\left[
 (g-1)b^{1-z}-(1-bg)
\right]
\\&=&(g-1)(b-1)\,(\log(b))^2b^{1-z}\,
\frac{ (g-1)b^{1-z}-(1-bg)}{\left((g-1)b^{1-z}+(1-bg)\right)^3}.
\label{unimodal yes}
\end{eqnarray}
This derivative can be zero (for $g>1$, $b>0$, $b\neq 1$)
if and only if the last ratio is zero.

\medskip

{\it (a) Case $bg < 1$.} This implies $1-bg>0$ and $b<1$,
and the term in front of the ratio in \eqref{unimodal yes} is negative
and the denominator in the ratio is positive.
In this case, the derivative is thus positive
if $0~<~b^{1-z}~<~(1-bg)/(g-1)$ and negative if $0<(1-bg)/(g-1)<b^{1-z}$.
Since $b^{1-z}$ is increasing in $z$, we have a monotonically increasing
density on $[0,1)$ if $(1-bg)/(g-1)\ge 1$ and we have a monotonically
decreasing density on $[0,1)$ if $(1-bg)/(g-1)\le b$.
For $b<(1-bg)/(g-1)<1$, we have a unimodal density
with critical point
\begin{equation*}
z^*=1-\frac{\log\left((1-bg)/(g-1)\right)}{\log(b)}~\in~(0,1).
 \end{equation*}

{\it (b) Case $bg=1$.} This implies $b<1$. The density
$f_{b,g}(z)=-\log(b)b^z$ is monotonically decreasing in $z$.
 
\medskip 
 
 {\it (c) Case $bg > 1$.} This implies $bg-1>0$. which means that the numerator of the last ratio in \eqref{unimodal yes} is thus always positive. We therefore need to analyze for $b\neq 1$ the ratio
 \begin{equation}
 \label{2nd equation}
\frac{b-1}{(g-1)b^{1-z}+(1-bg)}, 
\end{equation}
in order to determine the sign of the derivative $f'_{b,g}$.

\medskip 

{\it (c1)} Consider the first case $b<1$. In this case, we have a negative numerator in \eqref{2nd equation} and 
$b^{1-z}$ is increasing in $z\in[0,1]$. If 
$0<b^{1-z}<(bg-1)/(g-1)$, we have a positive derivative,
and for $0<(bg-1)/(g-1)<b^{1-z}$, we have a negative derivative.
Since in this case, $(bg-1)/(g-1)<b < b^{1-z}$ holds for all $z \in (0,1]$, we have a monotonically decreasing
density.

\medskip

{\it (c2)} Consider the second case $b>1$. In this case, we have a positive numerator in \eqref{2nd equation} and 
$b^{1-z}$ is decreasing in $z\in[0,1]$. If 
$0<b^{1-z}<(bg-1)/(g-1)$, we have a negative derivative,
and for $0<(bg-1)/(g-1)<b^{1-z}$, we have a positive derivative.
Since in this case, we have  $(bg-1)/(g-1)>b>b^{1-z}$ for all $z \in (0,1]$,
the density is decreasing.

\medskip
Finally, the case $b=1$ also immediately follows. This completes the proof.
\EndProof}

\bigskip

{\Beweis
{\bf Proof of Proposition \ref{prop bell shaped}.}
For $bg < 1$, we have the MBBEFD density
\begin{eqnarray*}
  f_{b,g}(z)
  &=& \frac{(g-1)(b-1)\log(b)b^{1-z}}{\left((g-1)b^{1-z}+(1-bg)\right)^2}
      ~=~  \frac{\frac{b(g-1)(b-1)\log(b)}{(1-bg)^2}b^{-z}}{\left(\frac{b(g-1)}{1-bg}b^{-z}+1\right)^2}
  \\&=&\frac{(b-1)\log(b)}{1-bg}\,
        \frac{\exp\left\{z\log(1/b)+\log\left(\frac{b(g-1)}{1-bg}\right)\right\}}{\left(\exp\left\{z\log(1/b)+\log\left(\frac{b(g-1)}{1-bg}\right)\right\}+1\right)^2}.      
\end{eqnarray*}
This proves the first claim. For the second claim, we remark that the function $t\mapsto \psi'(t)$ is symmetric
around the origin $t=0$. In view of our claim, there is $z=z^* \in (0,1)$ such that
$z \log(1/b)+\log(b(g-1)/(1-bg))=0$ if and only if $b<(1-bg)/(g-1)<1$.
Using reparametrization
\eqref{reparametrization 1} completes the proof.
\EndProof}

\bigskip

{\Beweis
{\bf Proof of Proposition \ref{prop exp curve 1}.}
First the function $G(z)$ is well-defined for all $z \in [0,1]$ due to the assumption that the function $b$ maps to the interval $(0,\infty)$ and that $b(0) \neq b(1)$. We compute the
 first and second derivative of $G$ and obtain
\begin{eqnarray*}
G'(0) &=&\frac{b'(0)}{b(0)}\, \frac{1}{\log(b(1))-\log(b(0))}~>~0,
\\
  G'(z) &=&\frac{b'(z)}{b(z)}\, \frac{1}{\log(b(1))-\log(b(0))}~\geq~0,
  \\
  G''(z) &=& \frac{b''(z)b(z)-b'(z)^2}{b(z)^2}\,\frac{1}{\log(b(1))-\log(b(0))}~\leq~0,         
\end{eqnarray*}
where the inequalities hold if and only if $b'(0) > 0$, $b'(z) \geq 0$ and $b''(z)b(z) - b'(z)^2 \leq 0$ for all $z \in [0,1]$, or $b'(0) < 0$, $b'(z) \leq 0$ and $b''(z)b(z) - b'(z)^2 \geq 0$ for all $z \in [0,1]$.  
 \EndProof}    

\bigskip

{\Beweis
{\bf Proof of Corollary \ref{cor exp curve 1}.} According to Theorem \ref{Exposure curves properties}, the function $F_Z$ defined in \eqref{distfct} satisfies for $z \in [0,1]$
\begin{equation*}
\begin{split}
 F_Z(z) &= \left(1 - \frac{G'(z)}{G'(0)}\right)\mathds{1}_{\{ z < 1\}} + \mathds{1}_{\{ z = 1\}} \\
 &= \left(1 - \frac{b'(z)}{b'(0)}\, \frac{b(0)}{b(z)}\right)\mathds{1}_{\{ z < 1\}} + \mathds{1}_{\{ z = 1\}}.
 \end{split}
\end{equation*}
The remaining statements then follow.
\bigskip
\EndProof}

{\Beweis
{\bf Proof of Proposition \ref{prop exp curve 2}.}
First the function $G(z)$ is well-defined for all $z \in [0,1]$ due to the assumption that $b(0) \neq b(1)$. We compute now the
 first and second derivative of $G$ and obtain
\begin{eqnarray*}
  G'(0) &=&\frac{e^{b(0)}}{e^{b(1)} - e^{b(0)}} \, b'(0)~>~0,
  \\
  G'(z) &=&\frac{e^{b(z)}}{e^{b(1)} - e^{b(0)}} \, b'(z)~\geq~0,
  \\
  G''(z) &=& \frac{e^{b(z)}}{e^{b(1)} - e^{b(0)}} [b'(z)^2 + b''(z)]~\leq~0,       
\end{eqnarray*}
where the inequalities hold if and only if $b'(0) > 0$, $b'(z) \geq 0$ and $b'(z)^2 + b''(z) \leq 0$ for all $z \in [0,1]$, or $b'(0) < 0$, $b'(z) \leq 0$ and $b'(z)^2 + b''(z) \geq 0$ for all $z \in [0,1]$.  
\bigskip
\EndProof}    

{\Beweis
{\bf Proof of Corollary \ref{cor exp curve 2}.} According to Theorem \ref{Exposure curves properties}, the function $F_Z$ defined in \eqref{distfct 2} satisfies for $z \in [0,1]$
\begin{equation*}
\begin{split}
 F_Z(z) &= \left(1 - \frac{G'(z)}{G'(0)}\right)\mathds{1}_{\{ z < 1\}} + \mathds{1}_{\{ z = 1\}} \\
 &= \left(1- \frac{e^{b(z)}}{e^{b(0)}} \frac{b'(z)}{b'(0)} \right) \mathds{1}_{\{z < 1\}} + \mathds{1}_{\{z = 1\}}.
 \end{split}
\end{equation*}
The remaining statements then follow.
\bigskip
\EndProof}  

{\Beweis
{\bf Proof of Proposition \ref{log family = exp family}.}
    Let $G$ be an exposure curve belonging to the logarithmic linked exposure family and write
    \begin{equation*}
        G(z) = \frac{\log(b(z))-\log(b(0))}{\log(b(1))-\log(b(0))},
    \end{equation*}
    for some twice differentiable function $b : [0,1] \to (0, \infty)$ fulfilling condition \eqref{log link cond 1} or \eqref{log link cond 2}. Then, by defining $m = \min\limits_{z \in [0,1]} \log(b(z)) - 1$ and $\tilde{b}(z) = \log(\log(b(z)) - m)$, we obtain for $z \in [0,1]$,
    \begin{equation*}
        \begin{split}
            G(z) &= \frac{\log(b(z)) -\log(b(0))}{\log(b(1)) -\log(b(0))} \\
            &= \frac{\left(\log(b(z)) - m\right) -(\log(b(0)) - m)}{(\log(b(1)) - m)-(\log(b(0)) - m)} \\
            &= \frac{\exp\left(\tilde{b}(z)\right)-\exp\left(\tilde{b}(0)\right)}{\exp\left(\tilde{b}(1)\right)-\exp\left(\tilde{b}(0)\right)}.
        \end{split}
    \end{equation*}
    This means that if the function $\tilde{b}$ fulfills condition \eqref{exp link cond 1} or \eqref{exp link cond 2}, then the exposure curve $G$ also belongs to the exponentially linked exposure family. The latter holds since for $z \in [0,1]$,
    \begin{eqnarray*}
          \tilde{b}'(z) &=& \frac{1}{\log(b(z)) - m} \, \frac{b'(z)}{b(z)}, \\
          \tilde{b}''(z) &=& \frac{\log(b(z)) - m}{\left[b(z) \left(\log(b(z)) - m \right) \right]^2} \, \left(b''(z)b(z) - b'(z)^2 \right)    - \frac{b'(z)^2}{\left[b(z) \left(\log(b(z)) - m \right) \right]^2},
    \end{eqnarray*}
    which implies
    \begin{equation*}
        \tilde{b}'(z)^2 + \tilde{b}''(z) = \frac{\log(b(z)) - m}{\left[b(z) \left(\log(b(z)) - m \right) \right]^2} \, \left(b''(z)b(z) - b'(z)^2 \right).
    \end{equation*}
    This means that for any $z \in [0,1]$,
    \begin{eqnarray*}
        \tilde{b}'(z)^2 + \tilde{b}''(z) \leq 0 &\iff \quad b''(z)b(z) - b'(z)^2 \leq 0, \\
        \tilde{b}'(z)^2 + \tilde{b}''(z) \geq 0 &\iff \quad b''(z)b(z) - b'(z)^2 \geq 0.
    \end{eqnarray*}
    This shows that any exposure curve belonging to the logarithmic linked exposure family belongs to the exponentially linked exposure family but either $b$ or $\tilde{b}$ may take a more complicated form. The other direction follows by using a similar argument.
\bigskip
\EndProof} 

{\Beweis
{\bf Proof of Lemma \ref{unimodal is in question}.}
The power logarithmic linked exposure example leads to a well-defined distribution due to \eqref{condition on a} and Propositon \ref{prop exp curve 1}. Set $y=(1-z/\alpha)^\delta \in  ((1-1/\alpha)^\delta,1]$ for $z\in [0,1)$.
The derivative \eqref{this is the derivative} is zero only if
\begin{equation*}
a^2(\delta-1)(\delta-2)-a(\delta-1)(\delta+4)y
        +2y^{2} =0.
      \end{equation*}
If there exist real-valued solutions, they are given by
\begin{eqnarray}\label{solutions square root}
  y_{\pm}  &=& a\, \frac{(\delta-1)(\delta+4) \pm \sqrt{(\delta-1)^2(\delta+4)^2
               - 8 (\delta-1)(\delta-2)}}{4}.
\end{eqnarray} 
We start with the case $\delta \in (1,2]$. In that case, we have
\begin{equation*}
  (\delta-1)(\delta+4) \le \sqrt{(\delta-1)^2(\delta+4)^2 - 8 (\delta-1)(\delta-2)}.
\end{equation*}
This implies for the smaller solution $y_-\le 0$, thus, this solution 
provides $y_- \not\in  ((1-1/\alpha)^\delta,1]$. For the bigger solution,
we have, using \eqref{side constraint a} in the second step,
\begin{equation}\label{bigger solution}
  y_+ \ge a\, \frac{(\delta-1)(\delta+4)}{4} > \frac{\delta+4}{4} >1 .
\end{equation}
Thus, also this second solution provides $y_+ \not\in  ((1-1/\alpha)^\delta,1]$, therefore
the density is monotone for $\delta \in (1,2]$.

Next, we analyze the case $\delta>2$. The term under the square root in \eqref{solutions square root}
is given by
\begin{eqnarray*}
(\delta-1)^2(\delta+4)^2
- 8 (\delta-1)(\delta-2)
&=&(\delta-1)\left[(\delta-1)(\delta+4)^2
    - 8 (\delta-2)\right]
  \\
&=&(\delta-1)\left[(\delta-1)(\delta^2+8\delta + 8)+8\right]~>~0.
\end{eqnarray*}
Thus, there are two real-valued solutions to \eqref{solutions square root}.
The bigger solution $y_+$ also provides \eqref{bigger solution}, and henceforth,
$y_+ \not\in  ((1-1/\alpha)^\delta,1]$. Therefore, we can focus on the smaller
solution $y_-$. It is given by
\begin{eqnarray*}
  y_{-}  &=& a(\delta-1)\left[ (\delta/4+1) - \sqrt{(\delta/4+1)^2
               -  \frac{\delta-2}{2\delta-2}}\right].
\end{eqnarray*} 
The square bracket is in $[0,0.1]$, therefore there are $\alpha >1$ and $a>1/(\delta-1)$
such that $y_- \in  ((1-1/\alpha)^\delta,1]$.
In particular, there is a critical point
$z^* \in (0,1]$ with $f_Z(z^*)=0$ in these cases and we easily see from \eqref{this is the derivative} that $z^{*}$ is a maximum. This does however not hold for any $\alpha >1$ and $a>1/(\delta-1)$.
\EndProof}

\bigskip

{\Beweis
      {\bf Proof of Lemma \ref{sine distribution and unimodal}.}
  In order to prove that the sine logarithmic linked exposure example leads to a well-defined distribution function, we need to show that the function $h$ defined through $h(z) = b''(z)b(z) - b'(z)^2$ satisfies $h(z) \leq 0$ for all $z \in [0,1]$ according to Proposition \ref{prop exp curve 1}. For this, it suffices to show that $h(0) \leq 0$ and $h'(z) \leq 0$ for any $0 \leq z \leq 1$. This indeed holds since
    \begin{equation*}
        \begin{split}
            h(0) &= b''(0) b(0) - b'(0)^2 \\
            &= - \alpha^2 \sin(\beta) \Big[ \sin(\beta) + a \Big]- \Big[ \alpha \cos(\beta)   \Big]^2 \\
            &= - \alpha^2 [1 + a \sin(\beta)]~<~0,
        \end{split}
    \end{equation*}
    where the last inequality is due to $a < - \frac{1}{\sin(\beta)}$. Furthermore, let $z \in [0,1]$, then
    \begin{equation*}
        \begin{split}
            h'(z) &= b'''(z) b(z) - b'(z) b''(z) \\
            &= - \alpha^3 \cos(\alpha z + \beta) \Big[\sin(\alpha z + \beta) +a \Big] - \alpha \cos(\alpha z + \beta) (-\alpha^2 \sin(\alpha z + \beta)) \\
            &= -  a \alpha^3 \cos(\alpha z + \beta) ~<~0,
        \end{split}
    \end{equation*}
    since $a > 0$, $\alpha > 0$ and $\alpha z + \beta \in (-\frac{\pi}{2}, \frac{\pi}{2})$ for any $z \in [0,1]$. From \eqref{sinusoidal derivative}, we see that this derivative has either no root or a unique root $z^* \in [0,1]$ satisfying
\begin{equation}
    \label{maximum sinusoidal}
    z^* = \frac{1}{\alpha} \left[\sin^{-1}\Big( \frac{a^2 -2}{a} \Big) - \beta \right],
\end{equation}
where the inverse function $\sin^{-1}(\cdot)$ is chosen to map to the interval $[- \frac{\pi}{2}, \frac{\pi}{2}]$. This root can only exist if $-2 \leq a \leq -1$ or $1 \leq a \leq 2$. Since $a$ was assumed to be positive, this means that the density $f_Z$ can only be unimodal when $1 \leq a \leq 2$. Note that this condition is not sufficient since we do not a priori obtain that there is a root $z^{*}$ lying in the interval $(0,1)$. However, in the case where this root $z^* \in (0,1)$ exists, it is clear from \eqref{sinusoidal derivative} that it corresponds to a maximum of the density.

\EndProof}

\bigskip

{\Beweis
      {\bf Proof of Lemma \ref{quadratic distribution and unimodal}.}
    The function $b$ of this example is strictly decreasing, which means that the necessary condition in order to obtain a well-defined distribution reads $b''(z) + b'(z)^2 \geq 0$ for all $z \in [0,1]$ according to Proposition \ref{prop exp curve 2}. This condition holds due to the assumption $\beta < - \sqrt{-2 \alpha}$. Furthermore, the roots of the derivative of the density in \eqref{quadratic derivative} are given by solving
\begin{equation*}
    2 \alpha z + \beta = 0 \quad \textrm{or} \quad (2 \alpha z + \beta)^2 + 6 \alpha = 0.
\end{equation*}
Note that by the conditions imposed on the parameters, we have $2 \alpha z + \beta < 0$ for all $z \in [0,1]$. This means that the only extrema in the interval $[0,1]$ can lie at
\begin{equation*}
    z_{-} = \frac{- \beta - \sqrt{-6 \alpha}}{2 \alpha} \quad \textrm{and} \quad z_{+} = \frac{- \beta + \sqrt{-6 \alpha}}{2 \alpha}.
\end{equation*}
Furthermore, since $\alpha < 0$ and $\beta < 0$, we obtain $z_{+} < 0$. This means that the density can have at most one extremum $z^*$ lying between 0 and 1, and we have that $z^{*} \in (0,1) \iff 0 > - \beta - \sqrt{-6 \alpha} > 2 \alpha$, which is equivalent to $- \sqrt{-6 \alpha} < \beta < -2 \alpha - \sqrt{-6 \alpha}$. Note finally that this extremum $z^{*} \in (0,1)$ corresponds to the maximum of the density due to \eqref{quadratic derivative}.

\EndProof}

\bigskip

{\Beweis
      {\bf Proof of Lemma \ref{power exp distribution and unimodal}.}
    Since the function $b$ of this example is strictly decreasing, the necessary condition in order to obtain a well-defined distribution reads $b''(z) + b'(z)^2 \geq 0$ for all $z \in [0,1]$ according to Proposition \ref{prop exp curve 2}. This condition holds due to the assumptions made on the different parameters. Indeed, these assumptions imply that $b'(z)$ is negative and decreasing, while $b''(z)$ is increasing for all $z \in [0,1]$. This means in particular that $b''(z) + b'(z)^2$ is increasing for all $z \in [0,1]$. In order to prove the above necessary condition, it thus suffices to show that
    \begin{equation*}
        b''(0) + b'(0)^2 > 0.
    \end{equation*}
    Given that $\alpha > 0$, $\delta > 0$ and $\epsilon < 0$, this is equivalent to
    \begin{equation*}
        \alpha (\alpha-1) \epsilon \delta^{\alpha -2} + (\alpha \epsilon \delta^{\alpha -1} - \beta)^2 > 0 \iff \left| \alpha \epsilon \delta^{\alpha -1} - \beta \right| > \sqrt{-\alpha (\alpha-1) \epsilon \delta^{\alpha -2}}.
    \end{equation*}
    By assumption, we have that $\beta > \alpha \epsilon \delta^{\alpha -1}$. This implies that the interior of the absolute value is negative, we can thus write
    \begin{equation*}
        \begin{split}
            b''(0) + b'(0)^2 > 0 &\iff  \beta - \alpha \epsilon \delta^{\alpha -1}  > \sqrt{-\alpha (\alpha-1) \epsilon \delta^{\alpha -2}} \\
            &\iff \beta > \alpha \epsilon \delta^{\alpha -1} + \sqrt{-\alpha (\alpha-1) \epsilon \delta^{\alpha -2}}.
        \end{split}
    \end{equation*}
    The parameter $\beta$ needs thus to satisfy precisely this last condition in order to obtain a well-defined distribution function.

\EndProof}

\bigskip

{\Beweis
{\bf Proof of Proposition \ref{Bernegger closed}.}
    Using \eqref{from exposure to dist}, the distribution of $\tilde{Z}$ can be written as
\begin{equation*}
    \begin{split}
    F_{\tilde{Z}}(z) &= \frac{F_Z(\bar{d}+z\bar{M}) - F_Z(\bar{d})}{1-F_Z(\bar{d})}
\,\mathds{1}_{\{z \in [0,1) \}} + \mathds{1}_{\{z =1\}} \\
&= \frac{G'(\bar{d})-G'(\bar{d}+z\bar{M})}{G'(\bar{d})}
\,\mathds{1}_{\{z \in [0,1) \}} + \mathds{1}_{\{z =1\}},
    \end{split}
\end{equation*}
for $z \in [0,1]$. By defining a new exposure curve $\tilde{G} : [0,1] \to [0,1], \, z \mapsto G(\bar{d}+z\bar{M})/G(\bar{d}+\bar{M})$ and using Theorem~\ref{Exposure curves properties}, we can derive the distribution function induced by $\tilde{G}$, which reads as
\begin{equation*}
    \begin{split}
        \tilde{F}(z) &= \left( 1- \frac{\tilde{G}'(z)}{\tilde{G}'(0)} \right) \mathds{1}_{\{z \in [0,1) \}}+\mathds{1}_{\{z =1\}} \\
        &=\left( 1- \frac{G'(\bar{d}+z\bar{M})}{G'(\bar{d})} \right) \mathds{1}_{\{z \in [0,1) \}}+\mathds{1}_{\{z =1\}}.
    \end{split}
\end{equation*}
This distribution is equal to the distribution of $\tilde{Z}$, which shows that $\tilde{Z}$ is a member of the Bernegger class with exposure curve $\tilde{G}$.
\bigskip
\EndProof} 

\newpage

\section{MLE parameters of the models of Section \ref{examples}}

\label{MLE parameters}

We provide in this appendix the values of all MLE parameters obtained when fitting the real dataset of Section \ref{examples} in the following tables.

\medskip

\begin{table}[htb!]
    \centering
    \begin{tabular}{l c c c}
      \toprule
        MLE parameters & $b$ & $g$ & $q$\\
      \midrule
      Standard MLE density (Green) & $5.66 \cdot 10^{-3}$& $50.5$ & - \\
      Flexible MLE density (Red) & $2.29 \cdot 10^{-3}$ & $94.9$ & $0.0339$ \\
      \bottomrule 
    \end{tabular}
    \caption{MLE parameters of the MBBEFD example rounded to three significant digits (Section~\ref{MBBEFD}).}
    \label{Tab: MBBEFD}
\end{table}

\begin{table}[htb!]
    \centering
    \begin{tabular}{l c c c c}
      \toprule
        MLE parameters & $\alpha$ & $\delta$ & $a$ & $q$\\
      \midrule
      Standard MLE density (Green) & $3.76 \cdot 10^4$ & $1.95 \cdot 10^{5}$ & $0.393$ & - \\
      Flexible MLE density (Red) & $9.32 \cdot 10^{3}$ & $5.67 \cdot 10^{4}$ & $0.275$ & $0.0339$ \\
      \bottomrule
    \end{tabular}
    \caption{MLE parameters of the power logarithmic linked exposure example rounded to three significant digits (Section \ref{power logarithmic linked}).}
    \label{Tab: Param Power log}
\end{table}

\begin{table}[htb!]
    \centering
    \begin{tabular}{l c c c c}
      \toprule
        MLE parameters & $\alpha$ & $\beta$ & $a$ & $q$\\
      \midrule
      Standard MLE density (Green) & $2.57$ & $-1.25$ & $1.05$ & - \\
      Flexible MLE density (Red) & $1.94 \cdot 10^{-3}$ & $-1.57$ & $1.00$ & $0.0339$ \\
      \bottomrule
    \end{tabular}
    \caption{MLE parameters of the sine logarithmic linked exposure example rounded to three significant digits (Section \ref{sine logarithmic linked}).}
    \label{Tab: Param Sine log}
\end{table}

\begin{table}[htb!]
    \centering
    \begin{tabular}{l c c c}
      \toprule
        MLE parameters & $\alpha$ & $\beta$ & $q$\\
      \midrule
      Standard MLE density (Green) & $-2.19$ & $-2.88$ & - \\
      Flexible MLE density (Red) & $-3.28$ & $-3.10$ & $0.0339$ \\
      \bottomrule
    \end{tabular}
    \caption{MLE parameters of the quadratic exponentially linked exposure example rounded to three significant digits (Section \ref{Quadratic exponential linked exposure example}).}
    \label{Tab: Param Quadratic log}
\end{table}

\begin{table}[htb!]
    \centering
    \begin{tabular}{l c c c c c}
      \toprule
        MLE parameters & $\alpha$ & $\beta$ & $\delta$ & $\epsilon$ & $q$\\
      \midrule
      Standard MLE density (Green) & $1.00$ & $-191$ & $0.130$ & $-195$ & -\\
      Flexible MLE density (Red) & $1.00$ & $-3000$ & $0.235$ & $-3000$ & $0.0339$\\
      \bottomrule
    \end{tabular}
    \caption{MLE parameters of the power exponentially linked exposure example rounded to three significant digits (Section \ref{power exponential link}).}
    \label{Tab: Param Power Exp}
\end{table}

\newpage

\begin{table}[t]
    \centering
    \begin{tabular}{l c c c}
      \toprule
        MLE parameters & $\mu$ & $\sigma$ & $\tilde{d}$\\
      \midrule
      MLE density & $-0.947$ & $0.563$ & $0.110$\\
      \bottomrule
    \end{tabular}
    \caption{MLE parameters of the lower-truncated and right-censored log-normal distribution rounded to three significant digits (Section \ref{comparison}).}
    \label{Tab: Comparison Lognormal}
\end{table}

\begin{table}[h!]
    \centering
    \begin{tabular}{l c c c}
      \toprule
        MLE parameters & $\gamma$ & $c$ & $\tilde{d}$\\
      \midrule
      MLE density & $1.97$ & $5.46$ & $0.0154$\\
      \bottomrule
    \end{tabular}
    \caption{MLE parameters of the lower-truncated and right-censored gamma distribution rounded to three significant digits (Section \ref{comparison}).}
    \label{Tab: Comparison Gamma}
\end{table}

\end{document}